\documentclass[epsfig,here,12pt,a4paper]{article}
\usepackage{epsfig} 
\usepackage{graphics}
\newcommand{\pa}{\partial}
\newcommand{\smallfrac}[2] {\mbox{$\frac{#1}{#2}$}}
\newtheorem{theorem}{Theorem} 

\renewcommand{\r}{\rightarrow}

\newcommand{\ggN}{g_{YM}^2 N}
\newcommand{\be}{\begin{equation}}
\newcommand{\ee}{\end{equation}}
\newcommand{\eel}[1]{\label{#1}\end{equation}}
\newcommand{\bea}{\begin{eqnarray}}
\newcommand{\eea}{\end{eqnarray}}
\newcommand{\eeal}[1]{\label{#1}\end{eqnarray}}
\newcommand{\baq}{\begin{equation}\begin{array}{rcl}}
\newcommand{\eaq}{\end{array}\end{equation}}
\newcommand{\eaql}[1]{\end{array}\label{#1}\end{equation}}
\newcommand{\beac}{\begin{equation}\begin{array}{rcl}}
\newcommand{\eeacn}[1]{\end{array}\label{#1}\end{equation}}
\newcommand{\ba}{\begin{array}}
\newcommand{\ea}{\end{array}}
\newcommand{\non}{\nonumber \\}
\newcommand{\equ}[1]{(\ref{#1})}
\renewcommand{\a}{\alpha}

\newcommand{\e}{\varepsilon}

 \newcommand{\al}{{\alpha^{'}}}
\newcommand{\beq}{\begin{eqnarray}} \newcommand{\eeq}{\end{eqnarray}}
\newcommand{\nn}{\nonumber} 
 \newcommand{\w}{Schwarzschild $\:$} 

\newcommand{\adss}{$AdS_5\times S^5$}

\newcommand{\pat}{\partial_t} 
\newcommand{\px}{\partial_x}
\newcommand{\crr}{\nonumber\\}       
\newcommand{\half}{{1\over 2}} 

\newcommand{\journal}[4]{{\rm #1~}{#2}\,(19#3)\,#4}

\newcommand{\pr}{\journal {Phys. Rev.}}

\newcommand{\np}{\journal {Nucl. Phys.}}

\newcommand{\gym}{g_{YM}}
\newcommand{\gef}{g_{eff}}
\newcommand{\preprint}[1]{\begin{table}[t]  
           \begin{flushright}               
           \begin{large}{#1}\end{large}     
           \end{flushright}                 
           \end{table}} 

\begin{document}
\begin{titlepage}

\preprint{TAUP-2623-2000 }

\vspace{2cm}

\begin{center}
{\bf  What does the string/gauge correspondence teach us about Wilson loops?}

\vspace{2cm}
{\bf  J. Sonnenschein}
\vspace{2cm}

{\em
 Raymond and Beverly Sackler Faculty of Exact Sciences\\
 School of Physics and Astronomy\\
 Tel Aviv University, Ramat Aviv, 69978, Israel}

\end{center}
\begin{abstract}
In these lectures we describe the 
attempt to extract the expectation values of Wilson loops   
from the string/gauge correspondence. We start with the original calculation
in \adss. It is then extended to the non-conformal
 background of $D_p$ in the near horizon limit. We discuss the computation 
at finite temperature. Supergravity models that admit confinement 
in 3d and 4d are described. A theorem  that determines
the classical values of loops associated with a generalized 
background is derived.
In particular we determine  sufficient conditions for confining behavior.
We then apply  the theorem  to various string models including type 0 ones.  
We describe   an attempt to 
 go beyond the classical string picture by incorporating  quadratic quantum 
fluctuations. In particular we address the BPS configuration of a single
quark, the supersymmetric determinant of \adss\  and   a setup  
that corresponds to a confining gauge
theory.
We determine the form of the Wilson loop for actions that include non trivial 
$B_{\mu\nu}$ field. The issue of an exact determination of the value of the
stringy Wilson loop is discussed. We end with a brief review of the baryons
from the string/gauge correspondence 

Lectures presented  in the ``Advanced School on Supersymmetry in the theories
of fields, strings and branes'' Santiago de Compostela-99.
\end{abstract}
\end{titlepage}
\newpage
\tableofcontents

\section{Introduction}
The idea to describe the Wilson loop of QCD in terms of a string partition
function dates back to the early Eighties.
For instance,  a landmark paper in this direction \cite{Luscher} showed that
the potential  of quark anti-quark separated at a distance $L$ acquires
 a  $-{c\over L}$ correction term ($c$ is a positive universal constant
independent of the coupling)  due to
 quantum fluctuations of a Nambu--Goto (NG) like action.
 An exact expression of  the partition function of the NG action
was  derived in the large $d$ limit \cite{Alvarez}, where $d$ is the
space-time dimension.
The result,  when translated to the quark  anti-quark potential,
took the  following form
$E(L)\sim {L\over \alpha'}\sqrt{1-{2c\over L^2/\alpha'}}$.
Thus, by  expanding it  for small  ${2c \over L^2/\alpha'}$,
one finds  the linear confinement potential  as well as the
so called L\"{u}scher term.  
It was later realized that  in fact this expression
is identical to the energy of the tachyonic  mode of the bosonic string
in flat spacetime with Dirichlet boundary conditions at $\pm L/2$
\cite{Arvis}.

Recently there has been a Renaissance of the
 idea  of a stringy description of the Wilson loop in the framework
of Maldacena's correspondence between large $N$ gauge theories and
string theory \cite{Mal,GKP,Wittenads,AGMOO}.
Technically, the main difference between the "old" calculations and the
modern ones is the fact that the spacetime background is no longer flat
but rather curved  and it includes additional ``relevant'' dimensions like
the  \adss or certain deformation  of it.
Conceptually, the modern  gauge/string  duality gave the stringy description
a more solid basis. The duality incorporate recipes  of extracting 
physical properties of the boundary theory like correlation functions
and others  from the dual string theory. The Wilson loop  falls into 
this category of physical gauge invariant properties that can be
read from the string picture.

The first "modern" computation \cite{Mal2, Rey} was 
a classical calculation based on  the \adss metric
which corresponds to the ${\cal N}=4$ supersymmetric theory.
The geometrical shape of the loop was taken to be that of a infinite 
strip. It was then generalized to the circle \cite{MalFish} and to 
closed loops with cusps \cite{DGO}. 
The latter  paper  further analyzed carefully 
the correspondence between the area of the string configuration and 
the Wilson loop as well as the contribution from the adjoint scalars 
to the expectation of the loop\cite{Gross}

To make contact with non-supersymmetric gauge dynamics one makes use of
Witten's idea \cite{Witads2} of putting the Euclidean time direction on a
circle
with anti-periodic boundary conditions.
This recipe was utilized to determine the behavior of the potential for
the  ${\cal N}=4$ theory at finite temperature \cite{RTY,BISY1}
 as well as 3d pure YM theory \cite{BISY2}
which is the limit of the former at infinite temperature.
Later, a similar procedure was invoked to compute Wilson loops 
related to  4d YM
theory,
 `t Hooft loops \cite{BISY2, GroOog} and  the quark anti-quark potential
in MQCD\cite{KSS1} and in Polyakov's type 0 model \cite{Polyakov,KleTse1}.
A unified scheme for all these models and variety of others was analyzed
in \cite{KSS2}. A theorem  that determines the leading and next to
leading behavior of the classical potential associated with this unified
setup was proven and applied to several models. In particular a corollary
of this theorem states the sufficient conditions  for the potential to have
a confining nature.

Another  map between string models and non-supersymmetric gauge theories 
is based on the correspondence with  type 0 string theories\cite{KleTse1}.
It was argued that the RR tachyon interaction shifts $m^2$ of the tachyons
to positive values. Area law Wilson loop characterizes the IR regime whereas
in the UV asymptotically free-like behavior was detected\cite{Minahan,KleTse2}.      

The baryonic configuration is another gauge invariant quantity that 
was identified in the string picture. First the baryonic vertex was
idntified\cite{witten}, then it was shown that the stringy baryons 
are stable\cite{BISY3} and finally they were related to exact solutions
of the BPS equations\cite{Baryon,Baryon1}

The issue of the quantum fluctuations and the detection of a L\"{u}scher
term
was raised again in the modern framework in \cite{GriOle}. It was noticed
there
that a more accurate evaluation  of the classical result \cite{KSS2}
did not have
 the form of a L\"{u}scher term. This is, of course, what one should have
anticipated, since after all  the  origin of the L\"{u}scher term \cite{Luscher}
is the quantum fluctuations of the NG like string.
The  determinant associated  with
the bosonic quantum fluctuations of the pure YM setup was  addressed
in \cite{GriOle}. It was shown
there that the system is approximately described by six operators
that correspond to massless bosons in flat spacetime and two additional
massive modes. The fermionic determinant was not computed in this paper.
However, the authors raised the possibility that the
latter will be of the form of massless   fermions   and hence there might
be a violation of the concavity behavior of gauge potentials \cite{Bachas,
DorPer}.In \cite{KSSW} it was argued  that in  fact the
fermionic operators are massive ones and thus the bosonic determinant
dominates and there is an attractive interaction after all.
The impact of the quantum fluctuations  for the case of the \adss case
was  discussed in \cite{Stefan, Naik}. Using the GS action
\cite{MetTse,Pesando,KalRah,KalTse}
with a particular $\kappa$ symmetry fixing,  it was observed that the
corresponding quantum Wilson loop suffers from UV logarithmic divergences.
It was argued that by renormalizing the mass of the quarks one can remove
the divergence.

The computations of Wilson loops in 3d and 4d pure YM theory  can be
confronted with the results found in lattice calculations. In  particular,
the main question is whether  the correction
to the linear potential   in the form of a L\"{u}scher term can be detected
in  lattice simulations.
According to \cite{Teper} there is some numerical evidence for a L\"{u}scher
term associated with a bosonic string, however the results are not
precise enough to be convincing.
Obviously,the ultimate dream is  compatibility with heavy meson
phenomenology.
The topics discussed in these lectures were addressed in varius other
publications\cite{other}

The  starting point of these lectures is the original determination of 
the Wilson loop\cite{Mal2}
of the ${\cal N}=4$  SYM theory  from the classical supargravity effective action
of the  type $II_B$ string theory on  the    \adss background. 
We then extend in section 2 this calculation also to the generalized
correspondence  between non-conformal SYM theory with 16 supersymmetries  
in $d+1$ dimensions and the supergravity solution  associated with  
$D_d$ branes in the the near horizon.  For that purpose we first briefly
describe this correspondence and then calculate the action of a static NG string
in those backgrounds.
In section 3 we analyze the behavior of the Wilson loop at finite temperature
using a near extremal supergravity solution. It is shown that there is a critical
separation length ( or critical temperature) above which the  quark anti-quark are 
fully screened.  
Section 4 is devoted to the supergravity background which corresponds to 
pure YM theory in  three  and (four) dimensions, or to be more precise the 
limit of $T\rightarrow \infty$ of  the maximal SYM theory in four ( five) 
dimensions. The set up is described,  area law behavior of the Wilson
loop  is found   as well as a screening behavior of 
a monopole anti-monopole pair via the 't Hooft loop.  
A generalized Nambu Goto action  for curved  backgrounds  with metric that 
depends on one coordinate  is written  down in section 5. A theorem 
that determines the leading and next to leading behavior of the Wilson loop
 associated with those scenarios is derived.
A corolarry  of that theorem determines  necessary conditions for having
an area law behavior.  We then present a table where  a list of models
are analyzed using the results of the theorem. 
Other string models which can be analyzed using the tools that follow
 from the theorem  are the various type zero models. 
In section 6 we briefly describe
 the string/gauge correspondence  for the non -supersymmetric cases
associated with type 0 models and then present the  calculations of the 
Wilson loops in the various energy scales. 
Section 7 is devoted to the impact of the string quadratic quantum fluctuations.
We address the issues of gauge fixing, the general structure of the 
bosonic determinant, the fluctuations in flat space-time, general
scaling relation,  the fermionic determinant in flat space-time, the determinant
of a single quark  in \adss and the Wilson loop  for that background and 
finally  the derivation of the  Luscher term  for the confining cases.
Section 8 is devoted to a discussion about possible exact determination
of the Wilson loop. 
In section 9  we briefly discuss the determination of the Wilson loops in
background with a non trivial WZ term.
Section 10 is devoted to  the analysis of baryons from supergravity.  The basic
 baryonic vertex is described and implemented  for the case of ${\cal N}=4$ 
 and for the more interesting confining models.

\section{ The Wilson  loop of ${\cal N}=4$  }

 The Wilson loop operator is 
$$ W({\cal C})={1\over N} Tr[Pe^{i\oint_{\cal C}A}]$$
where ${\cal C}$ is a closed loop in space-time.

\begin{figure}
\begin{center}
 \resizebox{8cm}{!}{
   \includegraphics{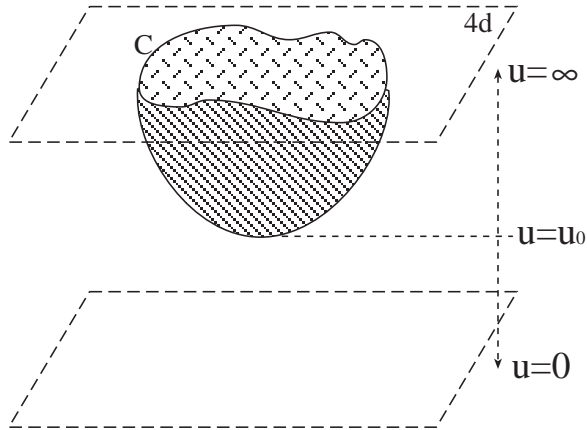}
   }
\end{center}
\caption{The basic setup of the Wilson loop}
\end{figure}

It is well known that one can extract the  potential energy $E(L)$ 
of a system of external  quark anti-quark  
 from $< W({\cal C})>$ via  
$$ < W({\cal C})>= A(L) e^{-TE(L)}$$
for a loop which is an infinite strip  ($T\rightarrow\infty$) (see fig. 2).
Here is these lectures we restrict our attention only to loops
with this type of shape. Loops of general shape, which are the natural objects
in discussions of the loop equation, and in particular  
circular loops\cite{MalFish} 
and loops with cusps were analyzed in \cite{DGO}.    

 In fact in ${\cal N}=4$ the adjoint scalars can also ``run'' along the loop
so the full Wilson loop is
$$ W({\cal C})={1\over N} Tr[Pe^{i\oint_{{\cal C}}ds  
A_\mu \dot\sigma^\mu + \theta^I(s)X^I(\sigma)\sqrt{\dot\sigma^2} }]$$
Here we restrict ourselves only to the gauge fields.  
 A way to introduce  the infinitely heavy external  quark pair is   terms of
  $W$ bosons associated
with the breaking of $U(N+1)\rightarrow U(N)$
where the expectation value of the Higgs field $\rightarrow \infty$

The natural candidate is
$$ < W({\cal C})>\sim e^{-S}$$
where $S$ is the proper area of the string world-sheet which on the boundary
of the $AdS_5$ is ${\cal C}$. 
\begin{figure}[h!]
\begin{center}
\resizebox{0.4\textwidth}{!}{\includegraphics{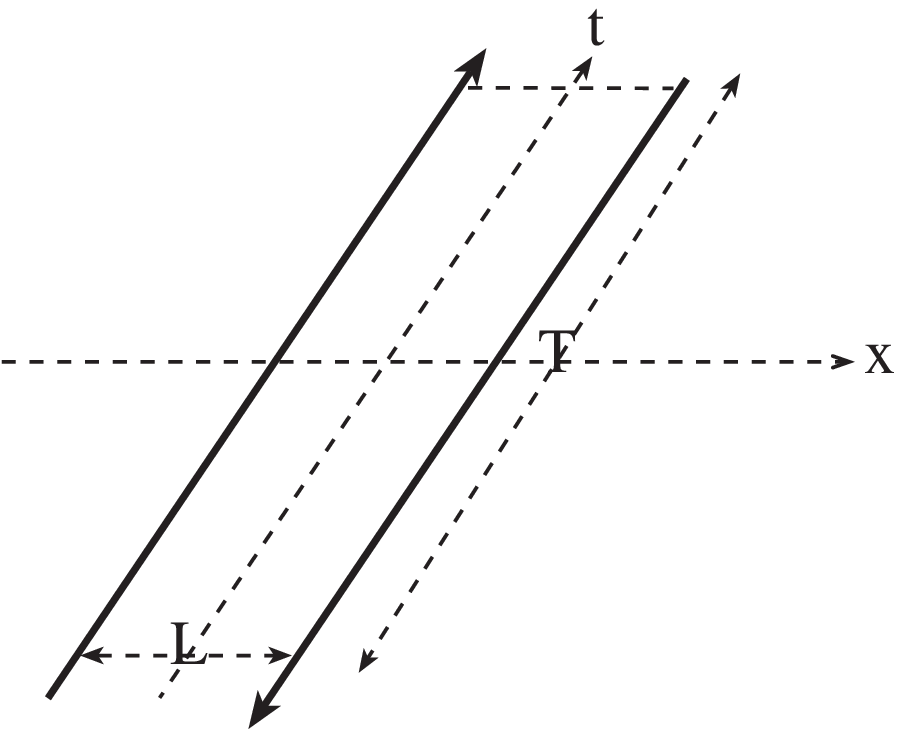}}
\end{center}
\caption{The Wilson loop setup for the \adss}
\end{figure}
 However, this includes the (infinite) mass of the W-bosons.
The subtracted expression is
$$ < W({\cal C})>\sim lim_{\Phi\rightarrow\infty} e^{-S_\Phi-l\Phi}$$
where $l$ is the length of the loop

In the framework of the correspondence between the   \adss supergravity
 theory and the large N, ${\cal} N=4$ super YM theory  $S$ is given 
in terms of  
 the Nambu-Goto action 
$$ S={1\over 2\pi\al}\int d\tau d\sigma\sqrt{det G_{\mu\nu}
\pa_\alpha X^\mu\pa_\beta X^\nu}$$
where $G_{\mu\nu}$ is the metric on the \adss  
\bea
&& ds^2  =  \alpha' \left\{ \frac{U^2}{R^2} 
[  dt^2 + dx_i^2] + R^2 
    \frac{dU^2}{U^2} + R^2 d\Omega_5^2 \right\} ,\non
&& R^2 = \sqrt{4 \pi g N}\nonumber
\eeal{ZX}
 Notice that as expected  the $\al$ factors disappear.

 Since we are interested in static configuration, 
it is natural to use the   ``static gauge"
$\tau=t$ and $\sigma=x$ . In this gauge  the action is  
\bea 
S &=& \frac{T}{2 \pi} \int dx \sqrt{(\partial_x U)^2 + (U^4/R^4} ~~.
\eeal{action}
 The ``Hamiltonian" in 
the $x$ direction is a constant of motion.
\be
\frac{U^4 }{\sqrt{(\partial_x U)^2 + (U^4 )/R^4}} = 
\mbox{const.}= R^2
\sqrt{U_0^4},
\ee This allows us to express $x$ as a
function of $U$ 
\be
x = \frac{R^2}{U_0}
 \int_1^{U/U_0}
\frac{dy}{y^2\sqrt{(y^4 - 1)}}
\ee
where $U_0$ is determined by  
\be
L  =  2\frac{R^2}{U_0}  \int_1^{\infty}
\frac{dy}{y^2\sqrt{(y^4 - 1)}}=  2\frac{R^2}{U_0}
\frac{\sqrt{2}\pi^{3/2}}{\Gamma(1/4)^2} 
\eel{lengthl}
 The expression for the energy is regularized by
integrating up to $U_{max}$ and  renormalized by
subtracting the mass of the 
$W$ boson  $U_{max}/2\pi$
This renormalization is equivalent for the \adss case to the 
use of the Legender transform of the energy advocated in \cite{DGO}.
The energy is given by 
\be
E = \frac{U_0}{\pi} \left\{  \int_1^\infty \left(\frac{y^2
      }{\sqrt{y^4 - 1}} - 1 \right) - 1 \right\}. 
\eel{energy} so that the final expression for $E(L)$is 
\be
E = -{2\sqrt{2}\pi^{3/2}(4\pi g^2_{YM}N)^{1/2}\over
\Gamma(1/4)^4}{1\over L} 
\ee
 The ${1\over L}$ dependence is dictated by conformal invariance 
since there is no  scale in the problem.
 Notice however the non-trivial dependence   $({g^2_{YM}N})^{1/2}$
which  differs from the perturbative YM result  $g^2_{YM}N$

\section{ Wilson loop  for the non-conformal  SYM theories
with 16 Supersymmetries} 
A correspondence, which is a  generalization   of the 
Ads/CFT  duality, between non-conformal  boundary field theories  in $p+1$
dimensions   
with 16 supersymmetries and the near horizon supergravity of $D_p$ branes 
 was conjectured  in \cite{IMSY}. 
The determination of the Wilson loop in those theories is thus 
 a straightforward generalization of the computation performed for the 
\adss background. We first briefly review the correspondence and then
describe the result for the quark anti-quark energy.

\subsection{A very brief review of the  generalized correspondence}
Let us examine now the generalization of the Ads/CFT duality 
to  a correspondence between the SYM theory with 16 supersymetries 
 in $p+1$ dimensions and the near horizon limit of the supergravity
background of $N$  Dp-branes\cite{IMSY}. 
Consider a  
system of $N$ coincident extremal  Dp-branes  
in the   following  limit \cite{Mal}  
\be\label{limit}
 \gym^2=(2\pi)^{p-2} g_s\al^{(p-3)/2}=\mbox{fixed},\;\;~~~~~
\al\r 0,~~~~~~U\equiv \frac{r}{\al}=\mbox{fixed},
\ee
where $g_s=e^{\phi_{\infty}}$, and $\gym $ is the  coupling
constant of the $p+1$ dimensional $U(N)$  SYM theory 
 that lives  on the $N$ Dp-branes. 
In the SYM picture $U$ corresponds to finite Higgs expectation value
associated with a  $U(N+1)\r U(N)\times U(1)$ symmetry breaking.
The effective coupling of the SYM theory is $\gef^2=\gym^2NU^{p-3}$.
Perturbation theory can be trusted in the region 
\be
\gef\ll 1.
\ee
The type II supergravity  solution describing $N$ coincident extremal
Dp-branes  is given by
\beq\label{bac}
&& ds^2=f_p^{-1/2}(-dt^2+dx_1^2+...+dx_p^2)+f_p^{1/2}
(dx_{p+1}^2+...+dx_9^2),\non
&& e^{-2(\phi-\phi_{\infty})}=f_p^{(p-3)/2},\\                
&& A_{0...p}=-\frac12 (f_p^{-1}-1)\nn,      
\eeq                                             
where $f_p$ is a harmonic function of the transverse coordinates 
$x_{p+1},...,x_9$
\be                                          
\alpha'^2 f_p= \alpha'^2 +\frac{d_p g^2_{YM} N }{U^{7-p}}~,~~~~~~~~~~~~~~
d_p = 2^{7-2p} \pi^{9-3p \over 2} \Gamma({7-p \over 2}) ~.
\ee 
In the special limit \ref{limit} the solution takes the form 
\beq\label{gsol}
&& ds^2=\al\left( \frac{U^{(7-p)/2}}{\gym\sqrt{d_p N}}dx^2_{||}+
\frac{\gym\sqrt{d_p N}}{U^{(7-p)/2}}dU^2+\gym\sqrt{d_p N}U^{(p-3)/2}
d\Omega^2_{8-p}\right) \non
&& e^{\phi}=(2\pi)^{p-2} 
\gym^2\left( \frac{\gym^2 d_p N}{U^{7-p}}\right) ^{\frac{3-p}{4}} \sim
 \frac{\gef^{(7-p)/2}}{N}.
\eeq

We will also consider near extremal configurations which correspond 
to the decoupled field theories at finite temperature. To do it we take
the limit (\ref{limit}) keeping the energy density on the brane finite. 
In this limit only the metric is modified and now reads

\beq \label{nearextr}
&&ds^2=\al\left\{ \frac{U^{(7-p)/2}}{\gym\sqrt{d_p N}}
[ - ( 1 -{ U^{(7-p)/2}_0 \over U^{(7-p)/2} } )dt^2 + dy^2_{||}]+
\right.\non
&& \left.
\frac{\gym\sqrt{d_p N}}{U^{(7-p)/2} ( 1 - {U^{(7-p)/2}_0 \over U^{(7-p)/2}} )
}dU^2+\gym\sqrt{d_p N}U^{(p-3)/2}
d\Omega^2_{8-p}\right\}
\eeq
The dilaton is the same as in (\ref{gsol}) and 
\be \label{enerdens}
U_0^{7-p} = { \Gamma({9-p \over 2} ) 2 ^{11-2p} \pi^{13-3p\over 2} 
\over (9-p) } \gym^4 \epsilon.
\ee
Here $\epsilon$ is the energy density of the brane above extremality and 
corresponds to 
the energy density of the Yang-Mills theory. With these formulas one
can calculate the entropy per unit volume 
and we find 
\be 
s = {S \over V} = \left({ \Gamma({9-p \over 2})^2 2^{43-7p} \pi^{13-3p} 
\over (7-p)^{7-p} (9-p)^{9-p} } \right)^{1 \over 2 (7-p)}
 \gym^{ p-3 \over 7-p} \sqrt{N} 
\epsilon^{ 9-p \over
2(7-p) } 
\ee
The temperature follows from the first law of thermodynamics.

These extremal and near extremal  solution 
can be trusted if the curvature in string 
units and the effective 
string coupling are small.
These conditions yield 
\beq
1 \ll  \gef^2 \ll N^{4 \over 7-p} ~ .
\eeq
which  translate for $p<3$ to the following range of the energy scale $U$
\beq\label{range} 
 (\ggN)^{1/(3-p)} N^{4/(p-3)(7-p)} \ll U\ll (\ggN)^{1/(3-p)} ~.
\eeq 
For $p>3$ the $\ll$ signs are replaced with $\gg$ ones. 
In the supergravity description $U$ is the radial coordinate.

From the limits on the SYM and supergravity regimes it is clear 
that  the `` phase diagram'' includes other domains apart from
these two.  The new domains occur when both the effective
coupling and the dilaton are large. 
When  the dilaton becomes large  one goes over to a description
in terms of M theory  for the even cases ( type $II_A$) whereas for
the odd $p$ cases ( type $II_B$) an s-duality maps the strong string
coupling to a weak coupling one.

A detailed description  of  all the regions is presented in  \cite{IMSY}.
The following figure summarize the  structure  of the theories 
along the  entire
$u$ axis  for the various $D_p$ branes.

\begin{figure}[h!]
\centerline{\psfig{figure=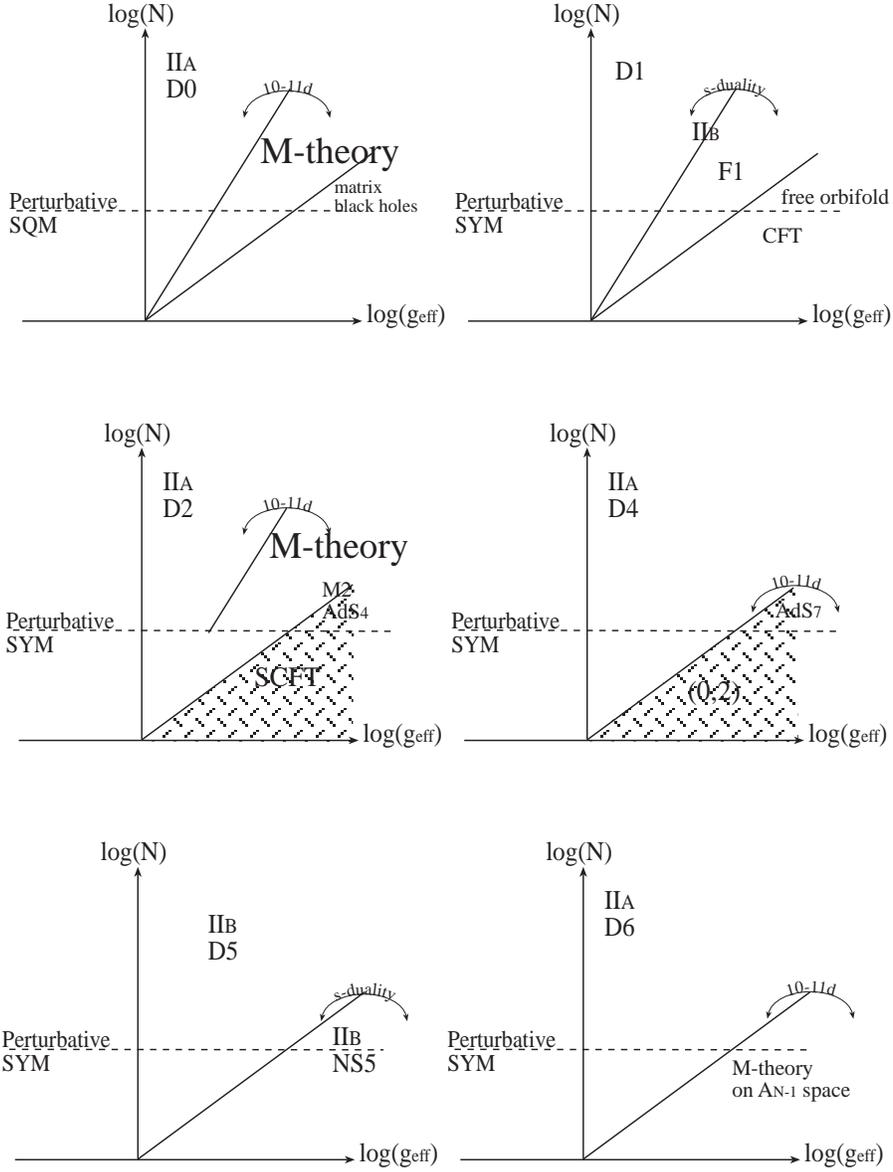,width=12cm,clip=}}
\caption{``Duality'' between SYM with 16 supersymmetries and the 
supergravity of $D_p$ in the near horizon limit..}
\label{fig:flow}
\end{figure}

\subsection{ The quark anti-quark energy }
By  applying  the procedure  used in the previous section 
for ${\cal N}=4$  to the  non-conformal, maximal 
supersymmetric   backgrounds 
one finds  
for the energy  of the quark anti-quark system   (for $p\ne5$)
\be\label{energyp}
E_{sg} \sim -({\ggN\over L^2})^{1/(5-p)},
\ee

This result  is valid 
as long as the  range of $U$ we integrate over  is in
the allowed supergravity region. This implies that 
(i) 
The minimal value of $U$, $U_0$,  has to be greater 
than the lower bound stated in \equ{range}, 
(ii) For $p<3$  since one cannot
integrate up to $U=\infty$,  to ensure a reasonable approximation
one has to demand that $x(U_{ub})-L/2\ll L$, where $U_{ub}$
is the upper bound of $U$.

 Generically   for $p<3$ 
the  result  can be trusted only for   the range \ref{range}
In certain cases  we showed that in fact the range
can be extended. 
\section{ Wilson loops at finite temperature}

 We want to study the finite temperature effects 
on the Wilson line deduced from the supergravity solution \cite{BISY1, RTY} 
We consider first  that system that 
corresponds to the ${\cal N}=4$ SYM in four dimensions.
 For  the finite temperature case  the relevant solution is the 
{\it near extremal solution} which is  Euclidean 
and periodic along the time direction.
The period  
is  the  inverse of the {\it  Hawking temperature}. 
The metric of the near extremal solution reads 
\bea
&& ds^2  =  \alpha' \left\{ \frac{U^2}{R^2} 
[ - f(U) dt^2 + dx_i^2] + R^2 f(U)^{-1}
    \frac{dU^2}{U^2} + R^2 d\Omega_5^2 \right\} ,\non
&& f(U)  =  1 - U^4_T/U^4, \\
&& R^2 = \sqrt{4 \pi g N} ,~~~~~~~~~~~ U_T^4 = 
\frac{2^7}{3} \pi^4 g^2 \mu ~,\nonumber
\eeal{metric}
where $\mu$ is the energy density above extremality on the brane and
the Hawking temperature  is $T =
U_T/(\pi R^2)$.
The action is now 
\bea 
S &=& \frac{T}{2 \pi} \int dx \sqrt{(\partial_x U)^2 + (U^4 -
U_T^4)/R^4} ~~.
\eeal{action}
  The ``Hamiltonian" in 
the $x$ direction is now 
\be
\frac{U^4 - U_T^4}{\sqrt{(\partial_x U)^2 + (U^4 - U_T^4)/R^4}} = 
\mbox{const.}= R^2
\sqrt{U_0^4 - U_T^4},
\ee
so that  $x$ as a
function of $U$ is 
\be
x = \frac{R^2}{U_0^3} \sqrt{U_0^4 - U_T^4}
 \int_1^{U/U_0}
\frac{dy}{\sqrt{(y^4 - 1)(y^4 - U_T^4/U_0^4)}} ~.
\ee
and in particular
\be
L  =  2\frac{R^2}{U_0} \sqrt{\e} \int_1^{\infty}
\frac{dy}{\sqrt{(y^4 - 1)(y^4 - 1 + \e)}} 
\eel{lengthe}
where $\e = 1 - U_T^4/U_0^4$.

 What about the subtraction? 
We subtract the (infinite) mass of the W-boson which 
corresponds to a string stretched between the brane at 
$U = \infty$ and the $N$ branes at 
 the 
horizon, $U = U_T$, and not at  $U = 0$. The
reasons for this subtraction are:
(i) 
It was shown that in the case of finite temperature
the coordinates of the supergravity solution are not identical to
the coordinates of the field theory living on the D-brane.
A coordinate transformation is needed to match them.
This transformation is such that from the point of view of
the field theory living on the brane (at the one-loop order)
the horizon is the origin.
(ii) The Euclidean solution  contains only the region outside the horizon. 
(iii) The local
 temperature close to the horizon is so  high
that it ``burns" any  static particle/string.
The Our last argument is due to Hawking radiation.
As is well known, due to the red shift-effect, the .
In fact it is so high that any static will burn.
In our case, by comparing the local temperature, $T_{loc}\sim
T_{Haw}\sqrt{g^{tt}}$ to the string mass, $M_s=1/\sqrt{\a'}$ 
The minimal distance for the string not to
burn is $U_T(1+1/R^2)\r U_T$ for $R\r \infty$.

Integrating from the horizon we obtain a finite result for the 
static energy 
\be
E = \frac{1}{\pi} \left\{ U_0 \int_1^\infty \left(\frac{\sqrt{y^4 - 1 +
      \e}}{\sqrt{y^4 - 1}} - 1 \right) - U_0 + U_T \right\}. 
\eel{energye}

\begin{figure}[h!]
\begin{tabular}{p{0.45\textwidth}p{0.45\textwidth}}
 \resizebox{0.45\textwidth}{!}{\includegraphics{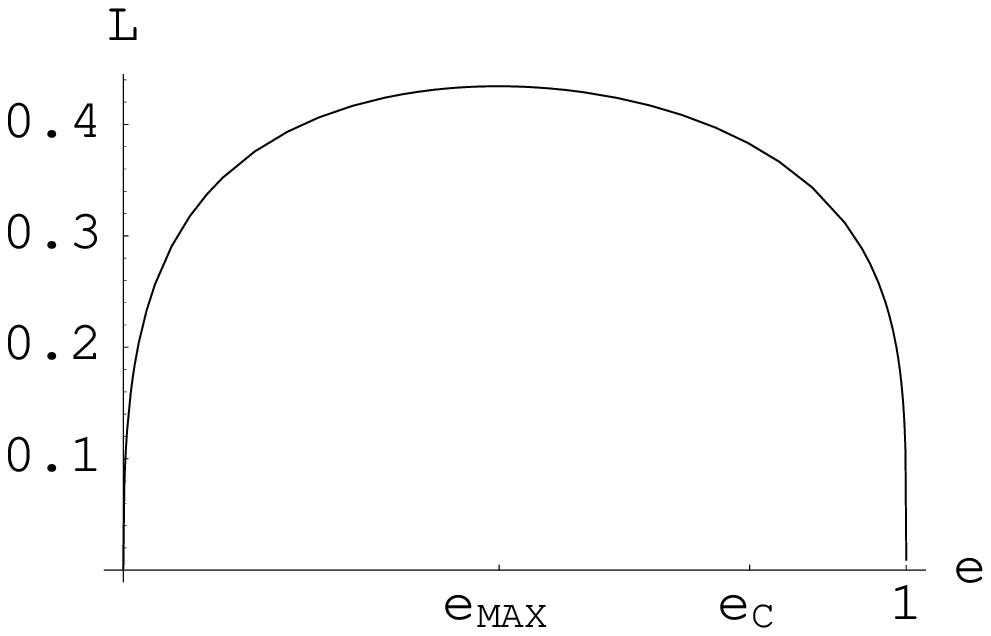}}
&
 \resizebox{0.45\textwidth}{!}{\includegraphics{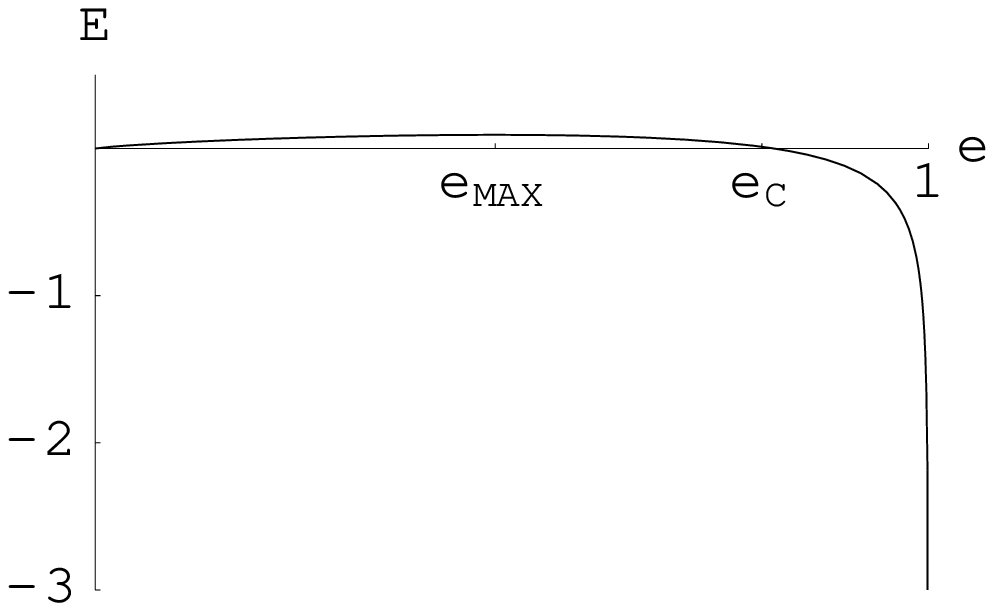}}
\\
\caption{The distance $L$ between the quark-anti-quark pair as a
  function of $\epsilon = \epsilon(T,\tilde U)$. Note that $L$ has a
  maximal value $L_{max}$.}
&
\caption{The energy of the quark-anti-quark pair relative to the
  ``free" quark situation  as a function of $\epsilon =
  \epsilon(T,\tilde U)$. Note that $E = 0$ is achieved at $\epsilon_C
  < \epsilon_{max}$ corresponding to $L_C < L_{max}$.}
\end{tabular}
\end{figure}
We can extract from it   $E(L,T)$    only numerically.
 At low temperature $TL\ll 1$ 
\be
E = -{2\sqrt{2}\pi^{3/2}(4\pi g^2_{YM}N)^{1/2}\over
\Gamma(1/4)^4}{1\over L}[1 + c(TL)^4] 
\ee
where $c$ is a positive numerical constant which does not depend on $R$.
The underlying conformal nature of the theory reveals itself in the 
fact that $E L$ can depend on $T$ only through the combination $T L$.
The behavior of $L$ as a function of $\epsilon$ seems a priori
puzzling since following   fig. 4
and fig. 5 
 $E(L,T)$ is double valued for $L > L_{max}$.

 Fortunately physics tells us
to believe the result only in the region $0<L<L_C$ where $L_C < L_{max}$.
 The existence of $L_C$ is seen in fig. 4. Starting from the low
temperature region $\epsilon \sim 1$ we reach $\epsilon_C$ at 
which $E = 0$. At this point the energy associated with our string
configuration (fig. 5) is the same as the energy of a pair of 
free quark and anti-quark pair. 

 Note  that $\epsilon_C$ is
reached before $L$ reaches its maximal value $L_{max}$ (fig. 4).

Once we reach $L_C$ our string configuration (fig. 5) does not
correspond to the lowest energy configuration which is $E=0$. 
 For a given temperature $T$ we encounter two regions with
different behavior. For $L << 1/T$ we observe a Coulomb like
behavior while for $L >> 1/T$ the quarks become free due to 
screening by the thermal bath.
In fig 8. we have plotted $E = E(L)$ for a given $T$ by eliminating
$\epsilon$ between eqs. \equ{lengthe} and \equ{energye} and trusting the
result up to $L_C$.

 Identical results are found for the correlation function 
of two temporal Wilson lines $W_t$ separated by a distance  $L$
$$ <W_t(0)W_t(L)>$$ 

\begin{figure}[h!]
\begin{center}
\begin{tabular}{p{0.3\textwidth}p{0.3\textwidth}}
 \resizebox{0.3\textwidth}{!}{\includegraphics{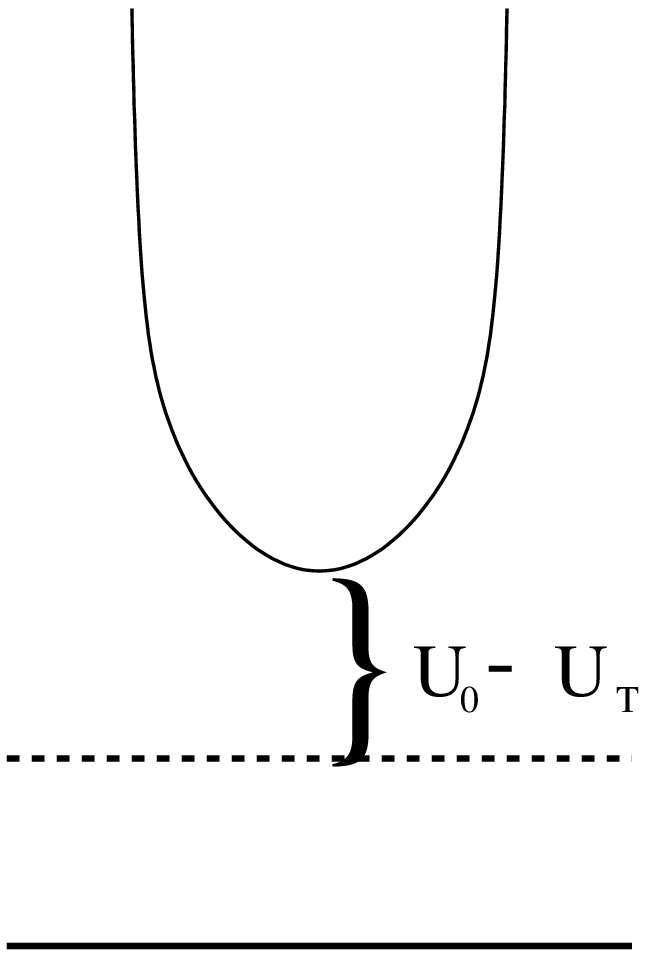}}
&
 \resizebox{0.3\textwidth}{!}{\includegraphics{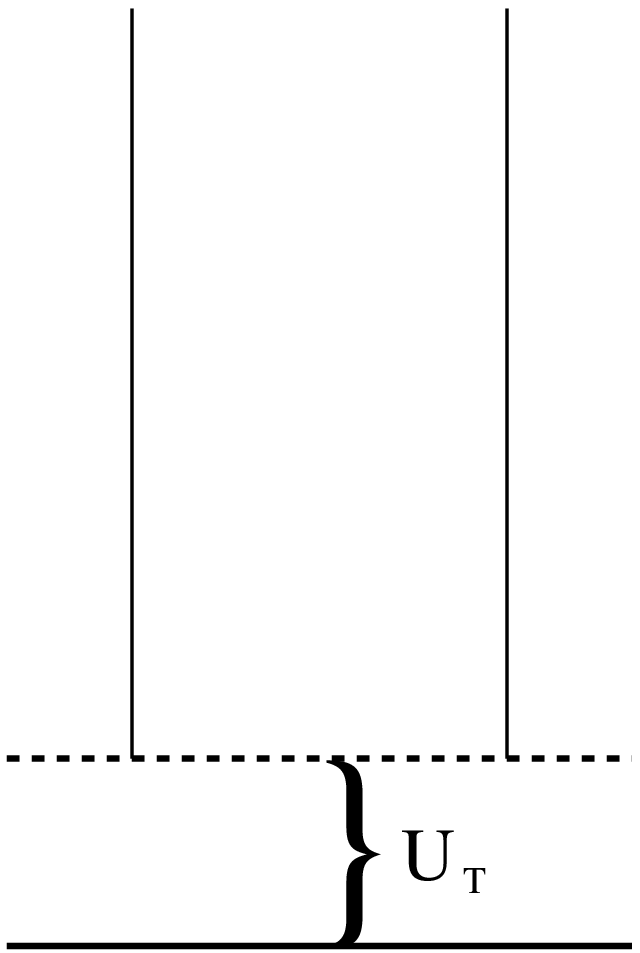}}
\\
\caption{The energetically favorable configuration for $L < L_C$.}
&
\caption{The energetically favorable configuration for $L > L_C$.}
\end{tabular}
\end{center}
\end{figure}
\begin{figure}[h!]
\begin{center}
\resizebox{0.4\textwidth}{!}{\includegraphics{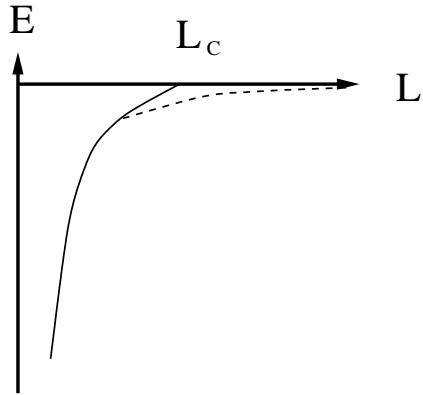}}
\end{center}
\caption{The energy $E$ of the quark-anti-quark pair as a function of
  $L$ for a given $T$. The solid line corresponds to the numerical
  calculation up to $L_C$, the dashed line indicates the expected
  behavior for large $L$.}
\end{figure}

\section{Area law behavior of Wilson lines from supergravity}
 Following the stringy interpretation of the Wilson loops  of
supersymmetric gauge theories at zero and finite temperature, 
 a natural question to ask
is whether one can  detect,
using the string/gauge duality,  the confining nature of pure YM theory.
We discuss here the computation in a setup that corresponds to ``pure YM" theory
in three dimensions, in four dimensional and finally we compute the 
't Hooft loop. 
\subsection{ Confinement in 3 dimensional YM theory}

The breaking of supersymmetry can be accomplished by choosing 
anti-periodic boundary conditions 
 on the compactified  time direction such that 
the ${\cal N}=4$ fermions and scalars acquire masses
$$m_{fermions}\sim T \qquad m_{scalars}\sim g_{YM4}^2 T$$
Decoupling of these fields occurs in the limit  of 
 $T\r \infty$ of the  Euclidean  4d  temperature. 
The effective theory in this limit is known to be   
 a  gauge theory in 3d with a coupling  
$$
g_{YM_3}^2 = g_{YM_4}^2 T ~.
$$
 The idea is, thus,  to 
consider the Wilson loop 
along two space directions for  the case of the 
near extremal D3 brane solution.


 The metric of near extremal D3 branes in the large N limit is 
\bea\label{metricYM}
&& ds^2  =  \a' \left\{ \frac{U^2}{R^2} 
[ - f(U) dt^2 + dx_i^2] + R^2 f(U)^{-1}
    \frac{dU^2}{U^2} + R^2 d\Omega_5^2 \right\} \non
&& f(U)  =  1 - U^4_T/U^4 \\
&& R^2 = \sqrt{4 \pi g N} ,~~~~~~~~~~~ U_T^4 = 
\frac{2^7}{3} \pi^4 g^2 \mu ~,\nonumber
\eea
where $\mu$ is the energy density.
and we take one direction,  $Y\r\infty$  and the other
direction, $L$, to be finite.
In this limit the wilson line is 
invariant under translation in the $Y$ direction.
 Therefore, for  $LT\gg 1$,
we obtain zero temperature non-supersymmetric
Yang-Mills theory in three dimensions.

Using  the \equ{metricYM} the relevant NG  action  for the spatial Wilson loop is
\be
S = \frac{Y}{2\pi}\int dx \sqrt{\frac{U^4}{R^4}+\frac{(\partial_x
    U)^2}{1-U_T^4/U^4}}.
\ee The distance between the quark and the anti-quark is
\be\label{jk}
L = 2 \frac{R^2}{U_0} \int_{1}^{\infty}
\frac{dy}{\sqrt{(y^4-1)(y^4-\lambda)}}, 
\ee
where $\lambda =U_T^4/U_0^4 <1$ and
 $U_0$ is the minimal value of $U$.
 The energy is 
\be\label{kj}
E=\frac{U_0}{\pi}\int_{1}^{\infty}dy\left( \frac{
    y^4}{\sqrt{(y^4-1)(y^4-\lambda)}}-1\right) +\frac{U_T-U_0}{\pi},
\ee
Notice that in the limit $U_0\r U_T$ ($\lambda\r 1$) we get $L\r\infty$. 
In this limit the main contribution to the integrals in \equ{jk}
and \equ{kj} comes from the region near $y=1$.
 Therefore, we get for large $L$
\be
E=T_{QCD} L.
\ee
The tension of the QCD string  is
\be\label{m}
T_{QCD} = \frac{\pi}{2} R^2 T^2,
\ee
Since both $T\rightarrow \infty$ and $R^2\rightarrow \infty$, it is clear
that in these limit one cannot get a finite string tension. The hope is
that the structure survives  also when one penetrate to the  small $R$ 
domain where obviously the supergravity approximation is not valid and one
has to incorporate the full string theory.

Note also that the area law behavior,
does not imply confinement in the 3+1 dimensional theory
with temperature.
 This result  interpolates between confinement in
2+1 dimensions ($L T \gg 1$) and Coulomb behavior in ${\cal N}=4$ YM
in 3+1 dimensions 
($L T \ll 1$).
Is it really a 3d pure  YM theory?  The answer is that it is {\bf Not}.
There are varius reasons for that\cite{BISY2}. 
One obvious one is that it is a construction of ${\cal N}=4$ in 4d 
at $T\rightarrow \infty$  which is in fact not identical to pure YM theory
in 3d. This manifest itself in the form of the meson meson interaction
\cite{MalFish, LoeSon} that is not dominated  by a glueball exchange.
A more general argument s that one cannot anticipate to deduce from 
gravity  higher spin states like those one find along a Regge-trajectory. 
Nevertheless, it seems that the deviations do not change the 
basic fact tat it admits confinement. 
 
\subsection{Area law in four dimensional YM theory}

The approach of the previous section to confinement
can be generalized to obtain confinement in four 
dimensions from supergravity.
We need to consider the supergravity solution
of near-extremal D4-brane in the decoupling limit.
A D4-brane is described in M-theory as a wrapped M5-brane so
from the point of view of M-theory we relate  the near-extremal
solution of M5-brane to confinement in four dimensions as was 
suggested in sec. 4 of \cite{witten}.

The near-extremal solution of D4-branes in the decoupling limit
is \cite{IMSY}
\beq\label{nearexD4}
&& ds^2=\a'\left[ \frac{U^{3/2}}{R_4^{3/2}}\left(
(1-U_T^3/U^3)dx_0^2 +dx_1^2+...+dx_4^2\right) +
\right. \non && \left.
 \frac{R_4^{3/2}}{U^{3/2}
(1-U_T^3/U^3)}dU^2 +R_4^{3/2}
\sqrt{U} d\Omega^2_4\right], \\
&& e^{\phi}=\frac{1}{(2\pi)^2} g_5^2 
\left( \frac{U^3}{R^3}\right) ^{1/4},\nonumber
\eeq
where $R_4^{3/2}=g_5 \sqrt{\frac{N}{4\pi}}$ and $g_5$ is the 5D
SYM coupling constant.

We would like to study  the spatial Wilson loop in the 
region where $L T \gg 1$. In this region the effective description
is via a non-supersymmetric YM theory in four dimensions
with coupling constant 
\be\label{54}
\gym^2 =g_5^2 T.
\ee
Unlike the supergravity solution which was used in the previous
section the supergravity solution \equ{nearexD4}, which we use here, cannot
be trusted for arbitrary $U$. $L$,  is related to $U$
by $L\sim 1/U$ and hence it is also bounded.

Before we perform the calculation of the spatial Wilson line
let us first find the upper bound on $L T$ and the bounds on 
 $\gym$ and $\gef^2
 = \gym^2 N$.
The restrictions on $U$ and hence on $U_T$,
are such that the curvature in string units and the effective 
string coupling are small. The result of these restrictions is \equ{range}
\cite{IMSY} 
\be
\frac{1}{N g_5^2}\ll U_T \ll \frac{N^{1/3}}{g_5^2}.
\ee
Therefore, the supergravity solution can be trusted only for 
distances
\be\label{LL}
N g_5^2\gg L\gg \frac{g_5^2}{N^{1/3}}.
\ee
To find the bound on $T$ we use the relation between $T$ and $U_T$
(the temperature can be obtained from  \equ{nearexD4} and $\left.T
=\frac{1}{4\pi}
\frac{dg_{tt}}{dU}\right|_{U=U_T}$)
\be\label{tem}
T=\frac{3\sqrt{\pi}}{g_5 \sqrt{N}}\sqrt{U_T},
\ee
to get
\be\label{T}
\frac{1}{N g_5^2}\ll T\ll \frac{1}{N^{1/3} g_5^2} ~.
\ee 
From \equ{54} we find that the four dimensional coupling constant is
bounded by
\be
\frac{1}{N} \ll \gym^2 \ll \frac{1}{N^{1/3}} ~.
\ee
We see, therefore, that in the large $N$ limit $\gym $ must go to zero.
For the four dimensional effective coupling, $\gef$, we have
\be
1\ll \gef^2 \ll N^{2/3}.
\ee
Thus the effective four dimensional 
coupling constant
$\gef$ must be large otherwise we cannot trust the 
supergravity description. Finally we turn to the bound for $T L$. 
To be able to  use the supergravity results described below we need 
to find a region where $T L\gg 1$. From \equ{LL} and \equ{T} we get
\be
N^{2/3}\gg TL\gg \frac{1}{N^{4/3}}.
\ee
Therefore, in the large $N$ limit there is a region for which 
we can trust our results. Note that unlike the 3d case, considered 
in the previous section, in the 4d case, for any finite $N$, $L$ is 
bounded.  

Let us now derive the area law behavior.
The action for the string in this case is
\be
S=\frac{Y}{2\pi}\int dx \sqrt{\frac{U^3}{R_4^3}+
\frac{\partial_x U^2}{1-U_T^3/U^3}},
\ee
Using the same manipulations as in \cite{Mal2} we get
\beq
&& L=2\frac{R_4^{3/2}}{U_0^{1/2}} \int_{1}^{\infty}
\frac{dy}{\sqrt{(y^3-1)(y^3-\lambda_4)}},\non
&& E=\frac{U_0}{\pi}\int_{1}^{\infty} dy \left( \frac{y^3}{\sqrt{
(y^3-1)(y^3-
\lambda_4)}} - 1 \right) + \frac{U_T - U_0}{\pi},
\eeq
where $\lambda_4=\frac{U_T^3}{U_0^3}$.
For $TL\gg 1$ we have $(U_T-U_0)/U_0\ll 1$ and the integrals are 
dominated by the region close to $y=1$.
Therefore, as in the previous section, we get
\be
E=T_{YM} L
\ee
where the string tension is 
\be 
T_{YM}=\frac{8}{27 \pi^2}\gym^2 N T^2 ~.
\ee
This agrees with known large $N$ results if $T$ is identified with
$\Lambda_{QCD}$ up to a $N$ independent constant factor.


\subsection{'t Hooft line}
In YM theories  the `` electric-magnetic" dual of the Wilson line is 
the 't Hooft line.  Just as one extract the quark anti-quark potential
from the former, the later determines the monopole anti-monopole potential.   
We would like to address now the question of how  to
 construct a supergravity model  which associates with the 't Hooft line.
 
 Consider  
the supergravity background  of near extremal $D_4$ branes \ref{nearexD4}.
 At large distances the 
effective theory is four dimensional along $x_1, x_2, x_3, x_4$.
The stringy realization of a monopole?
 is  a D2-brane ending on the
D4-brane. 
 The D2-brane is wrapped along $x_0$ so from the point of view 
of the four dimensional theory it is a point like object.
 The BI action of a D2-brane is 
\be
S=\frac{1}{(2\pi\a')^{3/2}}\int d\tau d\sigma_1 d\sigma_2 e^{-\phi}
\sqrt{\mbox{det} h},
\ee
where $h$ is the induced metric.
 For  the  D2-brane we consider,  which is infinite along one 
direction and winds the $x_0$
direction,  we get
$$
S=\frac{Y}{\gym^2}\int dx \sqrt{\partial_x U^2+ \frac{U^3-U_T^3}{R_4^3}},
$$
where we have used \equ{54}.
 Note the $1/\gym^2$ factor which is expected for a monopole.
The distance between the monopole and the anti-monopole is 
\be
L = 2 \frac{R_4^{3/2}}{U_0^{1/2}} \sqrt{\epsilon} 
 \int_1^{\infty} \frac{dy}{\sqrt{(y^{3} - 1)(y^{3}
    -1 + \epsilon)}} ~.
\ee 
where $\epsilon = 1 - (U_T/U_0)^3$.
The energy (after subtracting the energy corresponding to a free 
monopole and anti-monopole) is 
\be
E=\frac{2 U_0}{(2\pi)^{3/2}
\gym^2}\left[ \int_{1}^{\infty}dy\left( \frac{\sqrt{y^3
        -1+\epsilon}}{{\sqrt{y^3
        -1}}}\right) -1 \right]    +\frac{2(U_T-U_0)}{(2\pi)^{3/2}}.
\ee
 For $L T \gg 1$  
it is energetically favorable for the system to be in a configuration
of two parallel D2-branes ending on the horizon and wrapping $x_0$
So in
the "YM region"  we find 
{\it screening of the magnetic 
charge} which is another indication to confinement of the electric charge.

\section{ Classical Wilson loops - general results }
Equipped with the experience with the determination of the various loops
in sections 1-5, we would like now to  analyze the general setting
We introduce here a space-time metric that unifies all these models and determine
the corresponding static potential\cite{KSS2}. 

Consider a 10d space-time metric 
\be
 ds^2 = -G_{00}(s) dt^2 + G_{x_{||}x_{||}}(s) dx^2_{||} 
+ G_{ss}(s) ds^2
 + G_{x_{T}x_{T}}(s) dx^2_{T} 
\ee
where  
$x_{||}$ are $p$ space coordinates on a $D_p$ brane  and 
$s$ and  $x_T$ are the transverse coordinates. 
The corresponding Nambu-Goto action is 
$$
S_{NG} =\int d\sigma d\tau \sqrt{det[\pa_\alpha  x^\mu \pa_\beta x^\nu
G_{\mu\nu}]} \crr
$$
Upon using $\tau=t$ and  $\sigma=x$, where  $x$ is  one of the $x_{||}$
 coordinates,   the  action 
for a static configuration  reduces  to  
$$
S_{NG}= T \cdot \int dx\sqrt{ f^2(s(x)) + g^2(s(x)) (\pa_x s)^2} \crr
$$ 
where
\be
f^2(s(x))  \equiv  G_{00}(s(x))G_{x_{||}x_{||}}(s(x)) \qquad
g^2(s(x))  \equiv  G_{00}(s(x))G_{ss}(s(x)) 
\ee
and $T$ is the time interval. 
The equation of motion  (geodesic line) is 
$$
\label{sofx}
\frac{d s}{d x} = \pm \frac{f(s)}{g(s)} \cdot
\frac{\sqrt{f^2(s)-f^2(s_0)}}{f(s_0)}
$$
For a static string configuration connecting ``Quarks" separated  
by a distance $L$ we therefore have 
$$
\label{lgeneral}
L = \int dx = 
2 \int_{s_0}^{s_1} \frac{g(s)}{f(s)} \frac{f(s_0)}{\sqrt{f^2(s)-f^2(s_0)}} ds 
$$
The NG action   
and   the corresponding  energy  $E={S_{NG}\over T}$ are 
divergent.  
The action  is renormalized  by subtracting  the quark masses\cite{Mal2}.
 For the $AdS_5\times S^5$ case it is equivalent to the procedure 
suggested by \cite{DGO}.
\be
m_q = \int_0^{s_1} g(s) ds
\ee
So that the renormalized quark anti-quark potential is   
\be
E  =  f(s_0) \cdot L  
 + 2 \int_{s_0}^{s_1} \frac{g(s)}{f(s)}  
( \sqrt{f^2(s)-f^2(s_0)} - f(s)  ) ds - 2 \int_0^{s_0} g(s) ds 
\ee 
The behavior of the potential is determined by  the following 
theorem\cite{KSS2}. 
\begin{theorem}
\label{main}
Let $S_{NG}$ be  the NG action defined above, with functions $f(s),g(s)$ such
that:
\begin{enumerate}
\item \label{smoothf} $f(s)$ is analytic   for $0 < s < \infty$.
At $s = 0$, ( we take here that the minimum of $f$ is at $s=0$ ) its expansion is:
$$
\label{expansion}
f(s) = f(0) + a_k s^k + O(s^{k+1})
$$ 
with $k > 0 \;,\; a_k > 0$.
\item \label{smoothg} $g(s)$ is smooth for $0 < s < \infty$. At $s = 0$,
its expansion is:
$$
g(s) = b_j s^j + O(s^{j+1})
$$
with $j > -1 \;,\; b_j > 0$.
\item \label{positive} $f(s),g(s) \ge 0$ for $0 \le s < \infty$.
\item \label{increasing} $f'(s) > 0$ for $0 < s < \infty$.
\item \label{inftyint} $\int^\infty g(s)/f^2(s) ds < \infty$.

\end{enumerate}
Then for (large enough) $L$ there will be an even geodesic line asymptoting
from both sides to $s = \infty$, and $x = \pm L/2$.
The associated potential  is 
\begin{enumerate}
\item \label{conf}   if $f(0) > 0$, then
\begin{enumerate}
\item if $k = 2(j+1)$, 
\vskip 1cm
\begin{center}
\fbox{
$E = f(0) \cdot L -2 \kappa + 
                                    O((\log L)^\beta  e^{-\alpha L})$}
\end{center}
\item if $k > 2(j+1)$, 
\vskip 1cm
\begin{center}
\fbox{
$E = f(0) \cdot L -2 \kappa - d \cdot L^{-\frac{k+2(j+1)}{k-2(j+1)}} 
+ O(L^\gamma ) $.}
\end{center}
\end{enumerate} where $\gamma={-\frac{k+2(j+1)}{k-2(j+1)} - \frac{1}{k/2-j}}$
 and $\beta$,$\kappa$, $\alpha$ and   $d$  are
positive constants determined by the string configuration.

In particular, there is  {\bf linear confinement}

\item \label{noconf} if $f(0) = 0$, then if $k > j+1$,
\begin{center}
\fbox{
$E = -d' \cdot L^{-\frac{j+1}{k-j-1}}
 + O(L^{\gamma'})$} 
\end{center}
where $\gamma'={-\frac{j+1}{k-j-1} 
- \frac{2k-j-1}{(2k-j)(k-j-1)}}$ and 
  $d'$ is a  coefficient determined by the classical configuration.

In particular,   there is  {\bf no  confinement}
\end{enumerate}
\end{theorem}

A detailed proof of this theorem is given  in \cite{KSS2}.

\subsection{ Corollary- Sufficient conditions for confinement}
A corollary  that follows straightforwardly from the theorem
states  sufficient 
 conditions for confinement. In fact two sufficient conditions 
can be extracted. Confinement occurs 
if either of the two conditions is  obeyed:

 \begin{figure}[h!]
\begin{center}
 \resizebox{8cm}{!}{
   \includegraphics{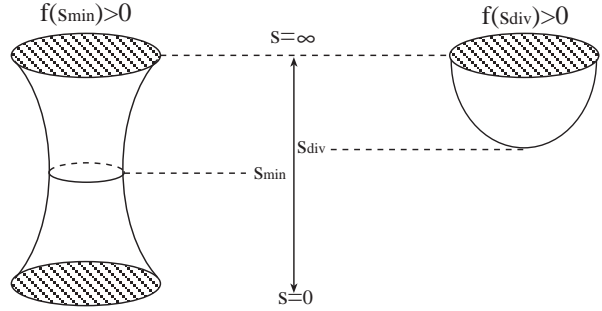}
   }
\end{center}
\caption{Sufficient conditions for confinement
On the left a space-time with a metric such that $f$ has a minimum
at $s_{min}$ with  $f(s_{min})>0$ and on the right a space-time with
horizon where $g$ diverges at $s_{div}$  and 
 $f(s_{div})\neq 0.$  
 }
\end{figure}

(i) {\bf $f$ has a minimum at $s_{min}$
 and $f(s_{min})\neq 0$}

or (ii) 
{\bf  $g$ diverges at $s_{div}$  and 
 $f(s_{div})\neq 0.$ }
\newpage

In the case that there are several minima to $f$ it is the one
with the greatest value of $s_{min}$ which is relevant. 
Models that obey the first condition $f(s_{min})\neq 0$ 
are characterized by having two disconnected boundaries (see fig. 9). 
The theory on  the boundary can be made unitary only provided that the
probability  of propagating a physical signal from one boundary component
to the other is negligible. Models that belong to the second condition
have generically a boundary theory that relate to a bulk theory with 
an horizon.

\subsection{ Applications to various models}

In the table below we summarize briefly   the application of the  theorem
to the models discussed in sections 2-4 as well as some additional 
models of the rotating brane, the $D_3-D_{-1}$ system and MQCD.

\begin{tabular}{|l||c|c|}\hline

Model& Nambu-Goto Lagrangian & Energy \\ \hline\hline
 & &    \\
$AdS_5\times S^5\cite{Mal2}$ &$ \sqrt{U^4/R^4 + (U')^2}$ &
$- \frac{2\sqrt{2}{\pi}^{3/2}R^2}{\Gamma(\frac{1}{4})^4}  \cdot L^{-1} $
 \\
 & &    \\\hline
 & &    \\
non-conformal & &    \\
$D_p$ brane\cite{IMSY,BISY2}   & $ \sqrt{(U/R)^{7-p} + (U')^2}$ & 
$ -d' \cdot L^{-2/(5-p)} $   \\
(16 susy)  & &  $+O(l^{-2/(5-p) - 2(6-p)/(5-p)(7-p)})$  \\
 & &    \\\hline
 & &    \\
Pure YM  in 4d at&
 & $\sim L^{-1}(1-c(LT)^4)$ for $L<<L_c$ 
  \\
 finite temperature &$ \sqrt{(U/R)^4 (1-(U_T/U)^4) + (U')^2}$ &
  full screening   $ L>L_c$\\
\cite{BISY2,RTY}& &   \\\hline
 & &    \\
Dual model of  & &   \\
Pure YM  in 3d &$  \sqrt{(U/R)^4 + (U')^2 (1-(U_T/U)^4)^{-1}}$ & 
$ \frac{U_T^2}{2\pi R^2} \cdot L -2\kappa + O((\log L)^\beta \; e^{-\alpha L})$   \\
\cite{Witads2,BISY2,DorOtt,GriOle} & &    \\\hline
 & &   \\
Dual model of  & &    \\
Pure YM  in 4d &$ \sqrt{(U/R)^3 + (U')^2 (1-(U_T/U)^3)^{-1}}$ &  
$\frac{U_T^{3/2}}{2\pi R^{3/2}} \cdot L -2\kappa + O((\log L)^\beta \; e^{-\alpha L})$
  \\
& &    \\\hline

 & &    \\
Rotating $D_3$ \cite{CORT} & $ \sqrt{C} \sqrt{ \frac{U^6}{U_0^4} \Delta + (U')^2 \frac{U^2 \Delta}
               {1-a^4/U^4 - U_0^6/U^6}}$ &$4/3{U_T^2\over R^2}C L+...$    \\
& &    \\\hline
 & &    \\
$D_3+D_{-1}$ \cite{LuiTse} &$ \sqrt{(U^4/R^4+q) + (U')^2(1+qR^4/U^4)}
$ & qL+...   \\              
 & &    \\\hline
 & &    \\
$MQCD$ \cite{KSS1} &$  2 \sqrt{2\zeta} \sqrt{\cosh(s/R_{11})} \sqrt{1+s'^2}$ &
$E = 2 \sqrt{2\zeta} \cdot L -2\kappa $ 
    \\  
& &  $+ O((\log L)^\beta \; e^{-1/\sqrt{2}R_{11} L})$  \\
 & &    \\\hline
 & &    \\
't Hooft loop 
&$\frac{1}{g^2_{YM}} \sqrt{(U/R)^3 (1-(U_T/U)^3)  + (U')^2 }$
&  full screening   \\
\cite{BISY1,GroOog} & & of monopole pair    \\\hline
\end{tabular}

\section{ Wilson loops associated with type 0 string models}
Three avenues were taken recently  on the way to extract information about 
non-superrsymmetric confining Wilson loops from  classical supergravity
effective actions. The first associated with the introduction of temperature
was discussed in section  4. The second  based on orbifolding of the 
\adss\cite{Kachru} is not   addressed in these lectures. The third 
option ( which is in fact equivalent to the second one) 
conjectures a duality between  boundary gauge theories and  type 0 string theories
 in the same spirit of the AdS/CFT correspondence. 
The determination of the potential
in such scenario is the topic of this chapter which we start with a brief
reminder of what are type 0 string theories.  

\subsection{ What is a type 0 string theory}
\begin{itemize}
\item Type 0 string is supersymmetric on the world sheet but not in space-time
due to a non-chiral GSO projection. 
In 10d the massless spectra of type $0_A$ and $0_B$ string theory is given by
\bea
type\  0_A\ & :& (NS-,NS-)\oplus  (NS+,NS+)\oplus  (R+,R-) \oplus  (R-,R+) \crr
type\  0_B\ & :& (NS-,NS-)\oplus  (NS+,NS+)\oplus  (R+,R+) \oplus  (R-,R-) 
\eeal{type0}
\item The $(NS-,NS-)$ states (which are projected out in   the  type II  theories)
are the {\it tachyons}.

 The $(NS+,NS+)$ are the space-time metric $G_{\mu\nu}$
the NS $B_{\mu\nu}$ and the dilaton $\phi$. 

The states  $(R+,R-) \oplus  (R-,R+)$ for a doublet of the vector $A_\mu$
and  $A_{\mu\nu\rho}$- the  three form of the type $II_A$ and similarly
$(R+,R+) \oplus  (R-,R-)$ is a doubling of the axion $\chi$, 
the RR $B_{\mu\nu}$ and the self dual four form  $C^+_{\mu\nu\rho\sigma}$.

\item Thus,  the   type $0_A$ and type $0_B$ differ from 
the  type $II_A$ and type $II_B$ in the following properties

(i) No space-time fermions from the closed string sector.

(ii) Doubling of the RR fields.

(iii) Tachyons.

\item The  type 0 effective actions  are consistent only provided
(i) The Tachyon $m_{tach}^2$ can be shifted to $m_{tach}^2>0$
(ii) No dilaton ( and possible other massless fields) tadpoles
(iii) Small string coupling and small curvature, 
$g_{st}<<1 \qquad {\cal R}<<1$.
It is plausible \cite{KleTse1,KleTse2} that due to the interaction with the 
RR fields the tachyons indeed loose their tachyonic nature. The absence of tadpoles
was shown in string tree level. Generically , condition (iii)   cannot  be obtained
throughout the whole range of scales  but only in part of it.

\item The low energy effective action of the type $0_B$ theories is composed of
 the  $(NS+,NS+)$ sector which is identical to the one of type $II_B$,
 a tacyon sector, the RR sectors and the coupling of the later with the 
tachyon.  The corresponding `` equations of motion" for $\phi, G_{\mu\nu},T$
and  $C_{\mu\nu\rho\sigma}$ take the following form
\bea
\phi &:& c_0 + 2\nabla^\mu\nabla_\mu \phi -4 \nabla^\mu\phi \nabla_\mu \phi 
- \half m^2 T^2=0\crr 
G_{\mu\nu} &:& R_{\mu\nu} + 2\nabla_\mu\nabla_\nu \phi 
-\smallfrac{1}{4}\nabla_\mu T  \nabla_\nu  T
-{e^2\phi\over 96} f_T(T)(F_{\mu\cdot} F_{\nu}^{\ \cdot} 
-\smallfrac{1}{10} G_{\mu\nu} 
F_\cdot F^\cdot )=0\crr 
T &:& (-\nabla^\mu\nabla_\mu + 2\nabla^\mu\phi\nabla_\mu  +m^2) T  
-{e^2\phi\over 240} f'_T(T)(F_{ \cdot} F^{ \cdot} =0\crr
 C_{\mu\nu\rho\sigma} &:& \nabla_\mu[f_T(T) F^{\mu\nu\rho\sigma\tau}]=0
\eeal{type0eqm}
where $c_0=-{D-10\over \alpha'}$ describes a non-critical string,  
$f_T(T)= 1+T+\half T^2 +...$ which may turn out to be $f_T(T)=e^T$.
and $m^2={D-2\over 4\alpha'}$. 

\item Since there is a doubling of the RR fields, one can have two  types of 
theories. One in which only one type of RR field is turned on and the 
other when both. The former is referred to as the {\it the electric } theory
or {\it magnetic} and the latter is the {\it dyonic } theory.
In the dyonic case the flux of  electric and magnetic RR field has to be 
equal. The interaction term between the tachyons and the RR fields is
characterized by $h(T)=f(T), h(T)=f^{-1}(T)$ and 
$ h(T)=f(T) + f^{-1}$
for the electric, magnetic and dyonic cases respectively.  
\end{itemize}

\subsection{ The string/gauge correspondence and the Wilson loop}
Before describing possible solutions of  the equations,  we discuss now
 the corresponding boundary gauge theories. 
\begin{itemize}
\item The critical electric type $0_B$\cite{KleTse1}  theory with $N$ electric $D_3$ branes 
is conjectured to be  dual
 to a non-supersymemtric  $SU(N)$ gauge theory with  six scalars in  the 
adjoint. 

\item 
The dyonic theory on the other hand\cite{KleTse2} does include fermions due
to the open string between the electric and magnetic $D_3$ branes, 
thus constituting an $SU(N)\times SU(N)$ gauge theory with 6 adjoint
scalars and 4 bifundamental fermions in the representation $(N,\bar N)+
(\bar N, N)$. 
\item The five dimensional  non-critical theories\cite{AFS} correspond to a pure 
$SU(N)$ YM theory   in the electric case and an   $SU(N)\times SU(N)$
with fermions in  one $(N,\bar N) + (\bar N, N)$ 
representation

\item Solutions of the equations of motion were studied thoroughly  for the
critical case \cite{KleTse1,Minahan}. Here we want to demonstrate
the procedure of extracting the `` gauge theory properties" for the 
5d non-critical case\cite{AFS}. 
\item We search for solutions of the equations \ref{type0eqm}  which are
functions of the fifth coordinate only 
$\{ \phi, G_{\mu\nu}, T, F_{\cdot}\}(\rho)$. For the following
parametrization of the metric
$$dS^2 = e^{-4\phi+4\lambda} d\rho^2 + e^{2\lambda}d x_\mu^2$$
 the equations read 
\bea
&& \ddot{\Phi}
+ \smallfrac{1}{2}(5+\smallfrac{3}{16}T^2)e^{-4\Phi+8\lambda}
- \smallfrac{5}{4}Q^2 h(T)e^{-2\Phi+8\lambda}=0 \ \label{Phi} ,\\
&&\ddot{\lambda}
- \smallfrac{3}{4}Q^2 h(T)e^{-2\Phi+8\lambda}=0 \label{f} \ ,\\
&&\ddot{T}
+ \smallfrac{3}{4}Te^{-4\Phi+8\lambda}
- 2Q^2h'(T)e^{-2\Phi+8\lambda}=0 \ .\label{T}
\eea

\begin{figure}
\begin{center}
 \resizebox{8cm}{!}{
   \includegraphics{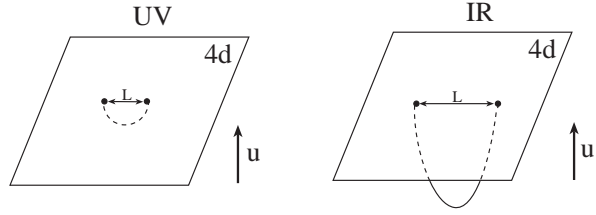}
   }
\end{center}
\caption{The IR and UV regimes-
For a small separation distance $L$ (left) the Wilson loop ``explores" only 
the large $u$ region and for large $L$ (right) it reaches small values of $U$}
\end{figure}

\item The dependence of the various fields on  the fifth coordinate $\rho$ ( which
was denoted in previous sections by $u$ or $s$) translates into the 
dependence  of the gauge dynamics on the scale- the renormalization group
flow.  To test such a correspondence one has to identify   what region
of the  fifth coordinate associates with the IR  and UV regimes 
in the gauge picture. Clearly, due to the possibility to transform
coordinates, the answer to this question should be based on   a ``meter" which
is a physical quantity. Nothing is more natural in the present
context than the use of  the Wilson loop itself as the measuring device.
From the   two   drawings  above it is clear that  
the $u\rightarrow \infty$ region corresponds  to small distance scales and hence to 
the  {\it UV regime}  and that the Wilson loop that stretch all the way to small
values of $u$ ends on the boundary on points separated with a large distance,
namely, it relates
to the {\it IR regime}.

\item From the equation of motion for the metric we find that $\ddot\lambda>0$\cite{AFS}
and hence the function $f$ defined in the theorem  also has $\ddot f>0$.
For boundary conditions such that $\dot\lambda<0$  the function $f$ has a minimum
at some $\rho_{min}$ at which 
$$f(\rho_{min})>0$$
Thus  the  solution obeys  the first sufficient condition for confinement that
was derived from the theorem of section 5. 
In fact one can show that this the structure of any generic solution, since the number
of parameters (four) in solutions found by expanding around $\rho_{min}$ matches the dimension
of the space of solutions.  
\item The conclusion is therefore that the generic solution to the equations of motion
admits an area law behavior in  the  IR regime.
Recall, however,  that to have a unitary theory on the boundary at $u\rightarrow\infty$
one has to assure that no physical signal can propagate from this boundary to the
other one at $\rho=0$. 

\item The confinement nature can also be verified using arguments based on the bulk 5d 
gauged-supergravity\cite{gppz} and in particular from the screening  nature of the
't Hooft loop. It is also important to note that in this regime one can trust
the supergravity solution since both  the string coupling  and the curvature are
small.
It is less clear to what extent a similar validity  characterizes that UV regime.
\item It was argued \cite{KleTse1} that there is a UV fixed point which locally 
takes the form of an \adss.  
Moreover around the fixed point $f$ behaves like a $log L$ so that
it was argued \cite {Minahan} that 
$$\Delta V_1\sim {1\over log{L_0\over L}}{1\over L}$$
namely, an asymptotic freedom behavior.
Klebanov and Tseytlin \cite{KleTse2}
 found  that the higher order correction 
produces a WIlson line  of the form 
$$\Delta V_2\sim {1\over log{L_0\over L}-clog\ log{L_0\over L}}{1\over L}$$
which resembles the 2 loop correction in  the gauge  theory picture.  
\end{itemize}

\section{Quantum fluctuations}
So far we have discussed Wilson loops from their correspondence to
certain classical string configuration. Now we write down  the machinery
to incorporate quantum fluctuations and present some  results
about the QM determinant of some of the classical setups discussed above
\cite{KSSW}.

We start with introducing   quantum fluctuations around the classical 
bosonic configuration
$$
x^\mu(\sigma,\tau) = x^\mu_{cl}(\sigma,\tau) + \xi^\mu(\sigma,\tau)
$$
The quantum correction  of the Wilson line is 
$$
        \langle W\rangle = {\rm e}^{-E_{cl}(L)T}~
        \int \prod_a d\xi_a \exp\left(-\int d^2 \sigma \sum_a \xi^a {\cal O}^a\xi^a \right)
$$
where $\xi^a$ are the fluctuations left after  gauge fixing. 
The corresponding  free energy is 
$$
F_B = -\log {\cal{Z}}_{(2)}= -\sum_a\half \log \det {\cal O}_a
$$

\subsection{ Gauge fixing}
In the classical treatment it is convenient to choose for the worldsheet
 coordinates $\tau=t$ and $\sigma=x$. 
In computing the quantum corrections it seems that there are several equivalent 
gauge fixings. One can still use the gauge of above, namely set $\xi_x=0$, or
fix $\sigma=u,\  \xi_u=0$ (we denote here $s$ by $u$) so that there are no
fluctuations in the  space-time metric. However, it turns out that those gauges
suffer from singularities at  the  minimum of the configuration $u_0$. 
A gauge that is free from those singularities is the ``normal coordinate
gauge" $\sigma=u_{cl}$ and the fluctuation 
in $x,u$ plane
is in the coordinate normal to  $u_{cl}$. 

An important subtle issue is whether associated with these gauge fixings
there are ghosts that have to
be taken into account.  As will be shown below it is clear that for 
the flat case there are none. The role of the ghosts for the
non-trivial backgrounds is less clear.  
\subsection{ General form of the bosonic determinant
}

In the $\sigma=u$ gauge and  after a change of variables the free energy 
is given by 
\be 
\label{FrEn}
       F_B = - \half \log {\det {\cal O}_{x}}
 - {(p-1)\over 2} \log {\det {\cal O}_{x_{II}}} 
- {(8-p)\over 2} \log {\det {\cal O}_{x_T}}
\ee
where 
\beq
\label {QuadOp} 
\hat{\cal O}_x &=& \left[\px\left((1-{f^2(u_0) \over f^2(u_{cl})})
\px\right) + {G_{xx}(u_{cl}) \over G_{tt}(u_{cl})} ({f^2(u_{cl}) \over f^2(u_0)}-1)
\pat^2 \right] \nonumber \\
\hat{\cal O}_{x_{II}} &=&  \left[\px\left({G_{y_i
y_i}(u_{cl}) \over G_{xx}(u_{cl})} \px\right) + {G_{y_i y_i}(u_{cl}) \over G_{tt}(u_{cl})} 
{f^2(u_{cl}) \over f^2(u_0)} \pat^2 \right]   \nonumber \\
\eeq
with $\hat{\cal O}= {2\over f(u_0)}{\cal O}$ and a similar expression
for $\hat{\cal O}_{x_T}$.
The boundary conditions are $\hat\xi(-L/2,t) = \hat\xi(L/2,t)=0$

\subsection{ Bosonic fluctuations in flat space-time }
Let us recall first the determinant in flat space-time.
The fluctuations  in this case are determined by the following action 
$$
        S_{(2)}=\half\int d\sigma d\tau
\sum_{i=1}^{D-2}\left[ 
        \left(\partial_\sigma\xi_i\right)^2 +
        \left(\partial_\tau\xi_i\right)^2
        \right]
$$
The  corresponding eigenvalues are 
$$
       E_{n,m} = ({{n \pi \over L}})^2 + ({{m \pi} \over T})^2
$$
and the free energy is given by
$$
       -\frac{2}{D-2} F_B =   \log \prod_{n m} E_{n,m} = 
 T {\pi \over 2L} \sum_n n + O(L)
$$

Regulating this result  using Riemann $\zeta$ 
function we find
that the  quantum correction to the linear quark anti-quark potential 
is

\begin{center}
\fbox{
$
      \Delta V(L) = - {1 \over T}  F_B 
= -(D-2) {\pi \over 24} \cdot {1 \over L}$
}
\end{center}

which is the so-called {\bf  L\"{u}scher term}\cite{Luscher}.

\subsection{General scaling  relation, and
 the $L$ dependence of $\Delta V$ 
}


Consider an operator of the form 
$$
{\cal O}[A,B] = A^2 F_t(v) \partial_t^2 + B^2 \partial_v ({\cal O}_v)
$$
where $v$ is a coordinate whose range is independent of $L$
The correction  to the potential $V[A,B]$
 due to fluctuations determined by such an
operator  is
 $$V[A,B] = (B/A) \cdot V[1,1]$$
For  the operators that describe the  fluctuations 
associated with  metrics such that
\be
f(u)  =  a u^k \qquad
g(u)  =  b u^j
\ee
like for  instance  for the $D_p$ brane solution in the near horizon limit
 we find that 
$A^2 = b u_0^j \;,\; B^2 = \frac{a^2}{b} u_0^{2k-j-2}\rightarrow 
B/A = \frac{a}{b} u_0^{k-j-1}$
Therefore, the
potential is proportional to 
\begin{center}
\fbox{
 $B/A = \frac{a}{b} u_0^{k-j-1}
\rightarrow \Delta V\propto L^{-1}$
}
\end{center}

Thus, the quantum correction of the quark anti-quark potential  is of  L\"{u}scher 
type\cite{Luscher}for models of $D_p$ branes with 16 supersymmetries, in particular also the 
$AdS_5\times S^5$ model.  
\subsection { The fermionic fluctuations in flat space-time }

The NSR action of the type II superstring with  RR fields like 
on $AdS_5\times S^5$ is not known.
On the other hand
the  manifestly  space-time  supersymmetric Green Schwarz  
action was written down for the $AdS_5\times S^5$ case\cite{MatTse}
 To demonstrate the use of the GS action we start with the 
fermionic determinant in flat space-time

 The fermionic part of the $\kappa$ gauged fixed  GS-action is
$$
    S_F^{flat} = 2 i  \int d\sigma d \tau  \bar \psi  \Gamma^i \pa_i \psi 
$$
where $\psi$ is a Weyl-Majorana spinor, $\Gamma^i$ are 
the SO(1,9) gamma matrices, $i,j = 1,2$ and we explicitly considered a flat classical string.
Thus the fermionic operator is
$$ 
    \hat{\cal O}_F=D_F = \Gamma^i \pa_i \qquad
(\hat{\cal O}_F)^2=\Delta = \px^2 - \pat^2$$  
The total free energy  is 
$$
    F = 8 \times \left(- \half \log \det \Delta + \log \det D_F \right) = 0 
$$
since for D=10, we have 8 transverse coordinates 
and 8 components of the unfixed  Weyl-Majorana spinor.
Thus in flat space-time the  energy associated with the supersymmetric string
is not corrected by quantum fluctuations.

\subsection{The determinant for a free BPS  quark of $AdS_5\times S^5$}

 
 The  $\kappa$ fixed GS action\cite{MatTse} is based on    treating
the target space as  the coset
 $SU(2,2|4) / (SO(1,4) \times SO(5))$.
 The  action incorporates the coupling to  the RR field. 
 The square of the  operator associated with the fermionic fluctuations 
of a BPS string representing a single quark is
$$
8\times \ \left( {\cal O}^2_\psi= ({R\over u})^2\pa_t^2 
+({u\over R})^2 \pa_u^2 -{7/4\over R^2} \right)
$$
where $R$ is the \adss radius.
 The bosonic operators are  of the form
\beq
3\times [ {\cal O}_x&=&       
({R\over u})^2\pa_t^2 
+({u\over R})^2 \pa_u^2 -{3\over R^2} 
\nonumber \\
5\times  [ {\cal O}_\theta&=&
({R\over u})^2\pa_t^2 
+({u\over R})^2 \pa_u^2 -{7/4\over R^2} ] \nonumber \\
\eeq
where $\{t, x, u,\theta\}\equiv 
\{\tau, {\xi\over u}, \sigma, \xi_\theta\}$
and $\theta$ is the coordinate on the $S^5$
 According to a theorem of McKean and Singer\cite{MacSin} the  divergences of
a Laplacian type operator of the form 
$$\Delta= \nabla^2+X  =-{1\over \sqrt{g}} D_a(g^{ab}\sqrt{g} D_b)  +X $$
vanishes if there is a match between the  fermionic and bosonic  coefficients of  
  $\nabla^2$  and $X$. In the present case  there are $8$ bosonic and $8$
fermionic $\nabla^2$ terms and hence there is no  quadratic divergences,
and the coefficients of $X$ are $8\times 7/4$ from the fermions and
$3\times 3+5\times 1$ from the bosons so  there are also no  logarithmic divergences. 
 It is thus clear that the divergent parts of the determinant
associated with the supersymmetric fluctuation of a BPS  string ``quark"
vanish. This problem is related to  issues associated with certain  
BPS 
soliton solutions\cite{JSV}.

\subsection{The  determinant for a Wilson line  of $AdS_5\times S^5$ }
 
 The GS action was further   simplified
by using a particular  $\kappa$ fixing\cite{KalTse}
\beq
S_{GS}&=&\int d^2\sigma  [\sqrt{g}g^{\alpha\beta} ( y^2[\pa_\alpha x^p
-2i\bar\psi\Gamma^p\pa_\alpha\psi][\pa_\beta x^p
-2i\bar\psi\Gamma^p\pa_\beta\psi]\crr
&+& {1\over y^2}\pa_\alpha y^t\pa_\beta y^t) +4\epsilon^{\alpha\beta}
\pa_\alpha y^t\bar\psi\Gamma^t\pa_\beta\psi ]\crr
\eeq
where $\psi$ is a Majorana-Weyl spinor
and the $Ads_5\times S^5$ metric is written in terms of the $4+6$
coordinates
$ ds^2 = y^2 dx_{II}^p dx_{II}^p +{1\over y^2} dy^i_{x_T}i dy^i_{x_T}$
The bosonic operators in the normal gauge now read 
\beq
2\times \qquad {\cal O}_{x_{II}}\ \ \ &=&\pa_x^2 
- {u^4\over u_0^4}\pa_t^2\nonumber \\
5\times \qquad {\cal O}_{\theta}\ \ \ &=&\pa_x^2 
- {u^4\over u_0^4}\pa_t^2 + 2{u^6\over u_0^4}\nonumber \\
1\times \qquad {\cal O}_{normal }&=&\pa_x^2 
- {u^4\over u_0^4}\pa_t^2 + 5u^2 -3{u^4\over u_0^2}\nonumber \\
\eeq
 The fermionic part of the action for the classical solution leads to the 
operator 
$$ 
    \hat{\cal O}^2_\psi = {u_0^2 \over R^2} \Gamma^1 \px + \left( {u_{cl}^4 \over {u_0^2 R^2}} 
\Gamma^0 + {u_{cl}^4 \over R^4} \cdot {\sqrt{u_{cl}^4-u_0^4} \over u_0^2} \Gamma^2 \right) \pat
$$
where we use $\Gamma$ matrices of SO(1,4), the  $AdS_5$ tangent space.
Squaring this operator, we find 
$$
    \left( {R^2 \over u_0^2} \hat{\cal O}_F \right)^2 = \px^2 - {u_{cl}^4 \over u_0^4} \pat^2  = {R^2 \over u_0^2} \hat{\cal O}_{II}
$$
Thus 
the transverse fluctuations ${\cal O}_{x_{II}}$ are cancelled 
by   the fermionic fluctuations. 
We are left with 6 fermionic degrees of
 freedom,  the normal bosonic fluctuation  and  5 additional bosonic
fluctuations associated with  
${\cal O}_{\theta}$.  
 Using our general result  we know that the quantum correction 
of the potential is  of a Luscher type. 
The universal coefficient and in particular  its sign  has not yet been determined.

In \cite{Stefan} the bosonic and fermionic determinants were 
analyzed in a different
gauge fixing procedure. 
The 2d reparametrization invariance was treated in the Riemann normal
coordinates and a diffent  fixing of the $\kappa$ symmetry was invoked.
Using these gauge fixings a closed expression for the supersymmetric
determinant was written down.
\be
W(C) = e^{-A_R}\;{\det\left(-\Delta_F - {1\over 4}R^{(2)} + 1\right)\over
\det\left(-\Delta_{cl}+2\right)
\det^{\half}\left(-\Delta_{cl} + 4 - R^{(2)}\right)
\det^{{5\over 2}}\left(-\Delta_{cl}\right)}
\ee
where  $A_R$ is the regulated area, $\Delta_{cl}$ is the classical 
Laplacian  and $\Delta_F=g^{\alpha\beta}\nabla_\alpha\nabla_\beta$
The corresponding free energy  suffers from 
logarithmic divergences. It was suggested to renormalize the quark masses
to get read of this divergence. 

\subsection{ The  determinant   for ``confining scenarios"}

Let us consider first the the setup which is dual to the pure YM theory
in 3d. For that case where the Euclidean coordinate is compactified 
\be
f(u)  =  u^2/R^2 \qquad g(u)  =  (1 - (\frac{u_T}{u})^4)^{-1/2}
\ee
so that $g$ diverges at the horizon $u=u_T$. In the large $L$ limit 
 \beq
\hat{\cal O}_{x_{II}} & \longrightarrow & \frac{u_T^2}{2R^2} \left[ \partial_x^2 +
  \partial_t^2 \right] \nonumber\\
\hat{\cal O}_t & \longrightarrow & {u_T^2\over 2R^2} e^{-2 u_T L} \left[ \partial_x^2 +
  \partial_t^2 \right] \nonumber\\
\hat{\cal O}_\theta & \longrightarrow & \left[ \frac{R^2}{2} [
   \partial_x^2 +  \partial_t^2 \right]
\eeq
 We see that the operators for transverse fluctuations, 
$\hat{\cal O}_{x_{II}}$, $\hat{\cal O}_t$, 
turn out to be simply the Laplacian in flat spacetime, multiplied by overall
factors, which are irrelevant.  Therefore, the
transverse fluctuations yield the
standard L\"uscher term  proportional to $1/L$. 
 The longitudinal normal fluctuations give rise to an operator 
$\hat{\cal O}_n$ corresponding to a scalar field with mass $2 u_T/R^2 = \alpha$.
Such a field contributes a Yukawa like 
 term $$\approx -\frac{\sqrt{\alpha}e^{-\alpha L}}{\sqrt{L}}$$
to the potential. 
Thus, in the metric that corresponds to  the 
``pure YM case"
there are 7 Luscher type modes and one massive mode.  
It can be shown that a similar behavior 
occurs in the general  confining setup \cite{KSSW}. 
Had the fermionic modes been those of flat space-time then the total coefficient
in front of the Luscher term would have been
a repulsive Culomb like potential\cite{GriOle} which  contradicts gauge
dynamics\cite{Bachas}. 
 However the point is that due to the RR flux 
the corresponding GS action  cannot be that of a flat space-time. Indeed
  the fermionic fluctuations  also become massive so that 
the  total interaction is attractive after all which is in accordance with 
\cite{DorOtt}.

\section{On the exact determination of stringy Wilson loops}
So far we have discussed the determination of Wilson loops from the classical
string  and the way  quantum fluctuations modify the classical result.
An interesting question to address is  whether in certain circumstances 
one can find an exact expression for the Wilson loop.  
Such an exact result was derived for the  simplest case of  
a string in flat space-time. Since the derivation was originally done  in terms
of the energy of the sting in the Polyakov formulation, we first show that
for a static case  the
latter is equal  to the NG action ( for a diagonal space-time metric).      

Recall  the form of the Polyakov   action in the covariant world sheet gauge
\be
S_{Poly}=T_{st}\int d^2\sigma \pa_\alpha X^\mu\pa^\alpha X^\nu G_{\mu\nu}(X)
\ee
and the  corresponding  space-time energy 
$$E=T_{st}\int_0^\pi d\sigma  \pa_\tau X^\mu G_{\mu 0}$$

Consider  a classical configuration that associates with a ``Wilson loop" 
 along the two coordinates $X(\sigma)$ ,
 $u(\sigma)$
and the rest of the coordinates are set to zero,  with
the following boundary conditions
$$X^1(0)=0; X^1(\pi)= L; \qquad u(0)=u(\pi)=\infty;\qquad X^0=\tau$$

Using  the equations of motion  it is a straightforward exercise to show that  

\be\label{energy}
 S_{NG}\sim E =\int_{u_0}^\infty {du\over\pa_xu }{f^2)\over f_0 }
\ee
where $x=X^1$, $f$ and $f_0$ were defined in section 5.

Consider now  the bosonic string in flat space-time
which implies that it stretches only along $x$
 with the boundary conditions of  above. 
\beq
 X^1(\sigma\tau)&=&{L\over \pi}\sigma + \sum_{n=1}^\infty 
{\alpha^i_n\over n}sin(n\sigma)e^{-in\tau} \nonumber \\
 X^0(\sigma\tau)&=& 2\pi\alpha'E\tau + i\sum_{n=1}^\infty {\alpha^0_n\over
n} cos(n\sigma)e^{-in\tau}  \nonumber \\
\eeq 
 The energy 
of the lowest {\bf tachyonic} state is given by
$$
E^2=P^iP_i + m_{tach}{^2}=({L^i\over 2\pi\alpha'})^2 -{(D-2)\over 24}{1\over
\alpha'} \nonumber\\
$$
so that\cite{Arvis} 
\begin{center}
\fbox
{$
E=T_{st} L\sqrt{1- {(D-2)\over 24}{1\over  T_{st}L^2}} 
 $}\end{center}
For $L>> ( T_s)^{-1/2}$ this can  be expanded to yield  
 $$\sim T_{st}L -\pi {(D-2)\over 24}{1\over L}+ ...
 $$
where the string tension $T^{-1}_{st}=2\pi\alpha'$.
Thus this expansion yields the Luscher  quadratic fluctuation term. 

 Moreover, this result is identical to the expression  for the
NG action   derived 
for a bosonic string in flat space-time  in the large
$D$ limit\cite{Alvarez}
$$ D\rightarrow \infty \qquad {\pi \over 24 T_{st}L^2}\rightarrow 0\qquad
{D\pi \over 24 T_{st}L^2}\rightarrow finite$$

A more challenging question is whether one can 
 find such ``exact" solution for a non-flat space-time.
A naive conjecture is that for the $Ads_5\times S^5$
the result is $\sim-{\sqrt{g^2N}\over L}\sqrt{1+{c\over \sqrt{g^2N}}}$.
However, whereas its large $g^2N$ expansion  
 includes the  result of \cite{Mal2} 
and a non trivial Luscher term, it does not permit a smooth extrapolation
to the weak coupling region where  the potential behaves like 
$\sim -{g^2N \over L}$. 
\section{ Wilson loops for string actions with  a $B_{\mu\nu}$ background}
  Exact results are known  for non trivial backgrounds  of
 group manifolds and coset spaces. That is one motivation to re-examine 
Wilson loops  in the presence of $B_{\mu\nu}$ background
which is an essential part of the action of a string on group manifold. 
In fact there are other independent reason to study this case. One additional
reason   is the $AdS_3$ string theory  that attracted recently a lot of attention and   
another one is the world-sheet approach to non-commutative geometry.

The bosonic action  in the presence of a WZ term can be written as 
$$S_B = S_{NG} + \int d^2 \sigma e^{\alpha\beta} \pa_\alpha X^\mu 
\pa_\beta X^\nu B_{\mu\nu}$$

For the application of using the string action to compute the Wilson loop
we consider a classical string configuration where $X^0(\tau)=\tau$
and the only non trivial coordinates are $X^1(\sigma)\equiv x(\sigma)$ and
$u(\sigma)$. For such configurations the only components of
$B_{\mu\nu}$ that can contribute are $B_{01}\equiv B$ and $B_{0u}
\equiv B_u$. We consider here the case that only  $B_{01}\equiv B$
 is non-trivial.
 In the coordinate
choice $\tau=x^0$ and $\sigma=x$ the WZ term takes the form 
$S_{WZs}= T\int dx [B(x)]$.  
Assuming
again a 
diagonal metric, the action in the static gauge takes the form
\be
 S_{NG+WZ} = T\int dx[\sqrt{f^2+ g^2 (\pa_xu)^2)}-B(x)]
\ee
were $f$ and $g$ were defined in section 5.
The equation of motion for this action takes the form
\be
{f^2\over \sqrt{f^2+ g^2 (\pa_xu)^2)}}-B = f_0-B_0
\ee
where $f_0$ and $B_0$ are the values of $f$ and $B$ respectively at $u_0$
where $u'=0$. For the case that $f=B$ having $u'=0$ at one point implies
$u'=0$ everywhere.
Moreover, in this case the action of the classical configuration vanishes.

 Consider  the case   where  $f\ne B$.
A particular simple case is when $B$ is a constant. 
Obviously for closed string the WZ term for such a case is a total derivative
and hence vanishes, however for the open string it is not. 
The  equations
of motion are not modified and the action is  just the  NG action 
plus an additional term of the form $S_B=TLB$. Note that such a correction
 translates
into an addition of a linear potential to the quark anti-quark interaction.

Using the boundary conditions the value of $u_0$ is related  to $L$ as
follows
\be
L/2 = \int_{u_0}^\infty {du}
={g\over f}{f_0+B-B_0 \over \sqrt{f^2-(f_0+B-B_0)^2}}
\ee
The action for a configuration that solves the equation of motion is
\be
S_{NG+WZ} =T\int_{u_0}^\infty {du}
{g\over f}{f^2-B(f_0+B-B_0)\over \sqrt{f^2-(f_0+B-B_0)^2}}
\ee
We can now determine the relation between the quark anti-quark potential  and
 the
separation disctance $L$ as follows
\be
E{q\bar q}=(f_0-B_0)L +2\int_{u_0}^\infty du
{g\over f} \sqrt{f^2-(f_0+B-B_0)^2}
\ee
since $E_{q\bar q}= {S_{NG}\over T}$

It is thus clear that for $f=B$ indeed the action and hence the energy vanish.
For sting  theoris where $f_0-B_0$ does not vanish  the Wilson loop
admits an area law behaviour  with  a string tension  equal to $f_0-B_0$.
This resutl holds provided that the second term is sub leading.
For the pure NG case ( with no WZ tern) this  was proven in the theorem
(see section 5)
 and the next to
leading classical  correction was found to be of the order $O(log
Le^{-\alpha L})$.

In the case with no $B$  the NG action was  renormalized by substructing
the ``masses of the two quarks'', namely,   
$
S_{Ren}= S-2\int_0^\infty du g(U)
$
(For a space-time metric that admits an horizon at $u=u_T$ the
range of integration starts from $u_T$ and not from $u=0$.)
A natural question is how to renomalize the $S_{NG+WZ}$ action.
It is obvious now that a straight line with $\pa_x u=\infty$ is not
a solution of the equation of motion. So one has to substruct 
the action associated with the quark string
$S_{Ren}= S-2 S^{quark}$

Let us  analyze now the Wilson loop of a string in 
an  $Ads_3$ background and in particular the $SL(2R)$ WZW model.
The action takes the form 
\be
{\cal L}_{SL(2R)} = \sqrt{(\pa_x u)^2+ u^4} -\alpha u^2
\ee
where 
 the  values of $\alpha$ are $\pm 1$ associated with the two possible
 orientation of the string  connecting
the points $x=0$ and $x=L$.
The equation of motion now reads
$$\alpha u^2 -{u^4\over\sqrt {(\pa_xu)^2 + u^4}}=(\alpha-1)u_0^2$$
where $u_0$ is a minimum of $u$ where  $(\pa_xu)=0$.

For $\alpha=1$ the solution is that $(\pa_xu)=0$ everywhere and thus the
 classical
configuration that  connects the two points on  boundary is just a straight line
 on
the boundary. This  is a  BPS string of zero energy.  
For $\alpha=-1$ again 
 one cannot connect two points at finite separation distance by
a string that penetrates the bulk.
As for the full exact quantum energy. It is clear that in the superstring case
 there
would not be any quantum corrections. However, for the bosonic string theory
one anticipates to have a Luscher term as the full answer of the following form
$ E={{3\over 1-2/k}+21\over 24}{1\over L}$
where  $k$ is   the level of the WZW model. 

It seems that  always in  a WZW theory with  aboundary  the value of
the action at the extremum vanishes. 
 In that case the only solution is
that $u'=0$ every where and so the two points on the boundary are
necessarily connected  by a straight line so the energy (which is zero )
is not a function of  $L$.
In a similar manner coset models based on a WZW theory with a boundary can
be analyzed.







\section{Baryon  Configurations from string ( supergravity) models} 

 Consider now on the AdS surface $N_c$ external quarks.
 What is the string system that can connect them?
Clearly  a string cannot connect a pair of quarks ( only a pair of a
quark anti-quark). 
 Alternatively the $N$ quarks can be connected to a {\it Baryon vertex}
The question is how can 
 we construct such a baryon vertex from the building blocks of the
type $II_B$ on the  backgroud?
 Witten \cite{witten} proposed 
 a baryon vertex  which  is based on a {\it a five brane wrapped on the
$S^5$} located at a certain point of the $AdS_5$.  
 The type $II_B$  self-dual five-form field  $G_5$
has   a flux of $N_c$ unites coming out the $S^5$. 
$\int_{S^5\times R}  {G_5\over 2\pi}= N$
On the $D_5$ world volume there is a coupling  of $G_5$ to an abelian
gauge field 
$$\int_{S^5\times R} a\wedge {G_5\over 2\pi}$$
So the $G_5$ contributes $N$ unites of $a$ charge. 
The total charge in the $S^5$ closed universe has to vanish.
 The strings  behave  as fermions so that the $N$ quarks
are in a completely antisymmetric representation

$N$ strings ending on the $D_5$ brane provide this needed charge


\subsection{  Baryons of ${\cal N}=4$ SYM  in four dimensions}

The baryons of the ${\cal N}=4$ SYM  theory are BPS states. 
The BPS equations associated with 
 the worldvolume Born-Infeld plus WZW
    action of a D5-brane in the background of N D3-branes
were studied in \cite{Baryon,Baryon1}.
 BPS-exact solutions were constructed. A   Hanany-Witten type 
mechanism was invoked   
 in the case that  a D5-brane is dragged across a stack of N D3-branes.
It was shown that  a
bundle of
N fundamental strings joining the two types of branes is created.
tes via the AdS/CFT correspondence.

Since we have in mind that our aim is the study of baryons in 
confining non-supersymmetric models, 
where the notion of a BPS state does not apply we
use here a different approach which is based on a ``zero order'' 
account of the action\cite{BISY3}. 
In such an approach we  consider independet 
contributions to the action from the
baryonic vertex and the strings, namely 
\be\label{action}
S_{total}=S_{D5}+N S_{1F}.
\ee

(i)The first contribution comes from the string stretched between the boundary  of the AdS$_5$ space
and the
D5-brane wrapped on the $S_5$.

(ii)  The second contribution comes from the D5-branes itself.
 a static D5-brane  wrapped on $S^5$ 
\be
S_{D5}=\frac{1}{(2\pi)^5 \al ^3 e^{\phi}}\int dx^6 \sqrt
{h}=\frac{T NU_0 }{8\pi},
\ee 
where $U_0$ is the location of the baryon vertex in the bulk, $T$ is
the time period which we take to infinity and 
 $h$ is the induced metric on the fivebrane.
\be
S_{1F}=\frac{T}{2\pi}\int dx \sqrt{U_x^2+U^4/R^4},
\ee
where $U_x=\frac{\partial U}{\partial x}$ and $R^4=4\pi g_s N$.
\begin{figure}
\begin{center}
 \resizebox{8cm}{!}{
   \includegraphics{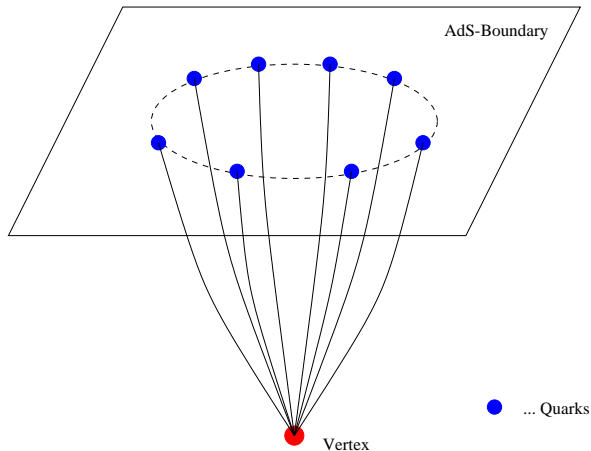}
   }
\end{center}
\caption{The Baryon Vertex}
\end{figure}
The configuration which we consider ( see Fig. 11)  is such that the strings end on a
surface with radius $L$ in a symmetric way.

The  stability conditions for a consistent baryonic configuration 
are the following:

(i) The symmetric quark configuration  ensures that 

$$\sum F_{x_i} =0$$ 

(where $x_i$ are the
direction along the boundary where the field theory is living.

(ii)  Along the $U$ direction the no-force condition is 

\be\label{2}
\delta U\frac{TN  }{8\pi} 
=\delta U\frac{TN(U_x)_0}{2\pi\sqrt{(U_x)_0^2+U_0^4/R^4}},
\ee
where $(U_x)_0= \frac{\partial U}{\partial x}|_{U_0}$
and $\delta U$ is the variation of $U$ at $x=0$ where the string
hits the baryon vertex. 

This condition can be derived as follows
The variation of (\ref{action}) under $U\rightarrow U +\delta U$
contains a volume term as well as a surface term.
The volume term leads to the Euler-Lagrange equation whose solution
satisfies \cite{IMSY} 
\be\label{1}
\frac{U^4}{\sqrt{U_x^2+U^4/R^4}}= \mbox{const.}
\ee
because the action does not depend explicitly on $x$.
The surface term yields the no force conditoin.

This   implies  the following relation between $U_0$ and the
radius of the  baryon $L$ 
\be
L = \frac{R^2}{U_0}  \int_1^{\infty}
\frac{dy}{y^2\sqrt{(\beta^2 y^4 - 1)}} ~
\eel{length}
where $\beta=\sqrt{16/15}$. 
The energy of a single string is given by 
\be\label{energyst}
E = \frac{1}{2\pi}U_0\left(  \int_1^\infty dy
\frac{\beta y^2}{\sqrt{\beta^2 y^4 - 1}} - 1 \right)
- \frac{U_0}{2\pi} 
\ee
Where we subtract the energy of the configuration with the D5-brane 
located at $U=0$.
Since  $g_{xx}$ vanishes at $U=0$  any radial string  which reaches
this point ends on the D5-brane.
As a result the energy of the fermionic strings, which we subtract
equals
  the energy of free
quarks.
Note that since $g_{tt}(U=0)=0$ the contribution  of the D5-brane located at
$U=0$ to the energy  vanishes.

Inserting the relation \equ{length} into \equ{energyst} 
one finds  that the  energy of each string is 
\be
E= -\a_{st} {\sqrt{2\gym^2 N  }\over L},~~\mbox{where}~~\a_{st}=~ 
\frac{1}{4} \sqrt{\frac{5}{6 \pi}} {}_2F_{1}[ \frac{1}{2}
, \frac{3}{4}, \frac{7}{4}; \frac{15}{16}] \times
{}_2F_{1}[\frac{1}{2}, - \frac{1}{4}, \frac{3}{4}; 
\frac{15}{16}]\simeq 0.036 
\eel{energytots}
The total energy of the baryon configuration is therefore
\be
E= -\a_B N {\sqrt{2\gym^2 N  }\over L},~~\mbox{where}~~\a_B=...\simeq 0.007 
\eel{energytotB}

Since the force $F=\frac{dE}{dL}$ is positive  the 
baryon configuration is stable.
 Moreover, as expected from 
the  field theory large N analysis, the energy  is proportional to $N$ 
times that of the quark anti-quark system.


\subsection{Baryons in non-SUSY theories}

We discuss YM in three dimensions (the generalization to the four
dimensional case is straight forward).
The supergravity solution associated with pure YM in three dimensions is
given by the near-extremal D3-branes solution in the decoupling limit
\be
ds^2 = \frac{U^2}{R^2} \left[ -(1 - \frac{U_T^4}{U^4}) dt^2 +
  \sum_{i=1}^3 dx_i^2 \right] + \frac{R^2}{U^2 (1 -
  \frac{U_T^4}{U^4})} dU^2 + R^2 d\Omega_5^2 ~. 
\ee
To obtain three dimensional theory we need to go to the IR limit and
to consider distances (along $x_1, x_2, x_3$) which are much larger
then $1/T$.
At the region where we can trust the supergravity solution, $R^2\geq
1$, the theory is not quite three dimensional 
YM theory.
Nevertheless, it does possess  the properties of YM in three
dimensions  which are relevant to the present discussion
\cite{Witads2,BISY2,Li,Oz}.

\begin{figure}
\begin{center}
 \resizebox{8cm}{!}{
   \includegraphics{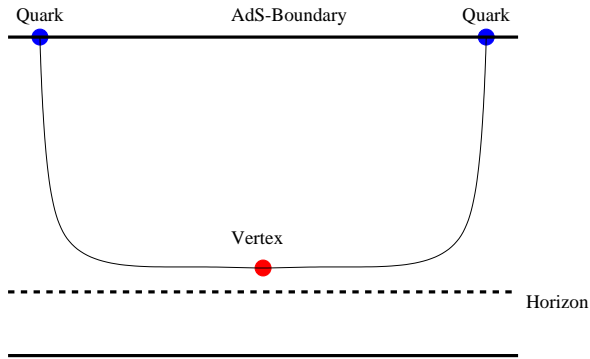}
   }
\end{center}
\caption{The baryon in non supersymmetric theories}
\end{figure}

The surface term gives 
\be\label{suoor}
\frac{1}{4}= \frac{U_0'}{(1-U_T^4/U_0^4)
\sqrt{U^4/R^4+(U_0')^2/ (1-U_T^4/U_0^4)}}.
\ee
To go to the IR limit we need to consider large $L$.
As in the Wilson loop case this means that $U_0\rightarrow U_T$.
At this limit the integrals of $L$ and $E$ are controlled by the region
near $U_0$ and their ratio is a constant which determins   
the QCD string tension.
At first sight it seems that eq.(\ref{suoor}) will change dramatically
the relation between $E$ and $L$.
However, at the IR  limit ($U_0\rightarrow
U_t$)  eq.(\ref{suoor}) implies that $U_0'$ 
 vanishes so the relation is again, as expected, linear
\be
E=N T_{YM} L, ~~~\mbox{where} ~~~
T_{YM} = \pi R^2 T^2.
\ee 

We should note the same relation holds for non-supersymmetric 
YM in four dimensions with the string tension found in  \cite{BISY2}.

\subsection{Baryons with $k<N$ quarks}
An important test to the stringy bayonic configurations 
of confining backgrounds is the question whether there are 
stable   baryons made out of  $k$ quarks when
$k<N$.
For example the case $k=N-1$ gives rise to a baryonic configuration in
the anti-fundamenatl representation.
In a confining theory we do not expect to find such a state (it cost
 an infinite
amount of energy to separate $N-k$ quarks  
all the way to infinity leaving behind the k-baryonic system).

Let us start with the setup that corresponds to the ${\cal N}=4$ SYM 
theory.  
The way supergravity enables us to construct  
baryons with less quarks is illustrated in figure 4.
In this figure we have
the usual baryonic vertex with $k$ strings stretched out to the
boundary at $U=\infty$ and the rest $N-k$ strings reaching $U=0$.
 
This configuration is stable provided that $\frac{dE}{dL}>0$.
The calculation of the energy of this configuration proceeds in a
similar way to the calculation of the energy of $k=N$ baryonic system
carries in section 2.
the surface term gives now the following relation

\begin{figure}
\begin{center}
 \resizebox{8cm}{!}{
   \includegraphics{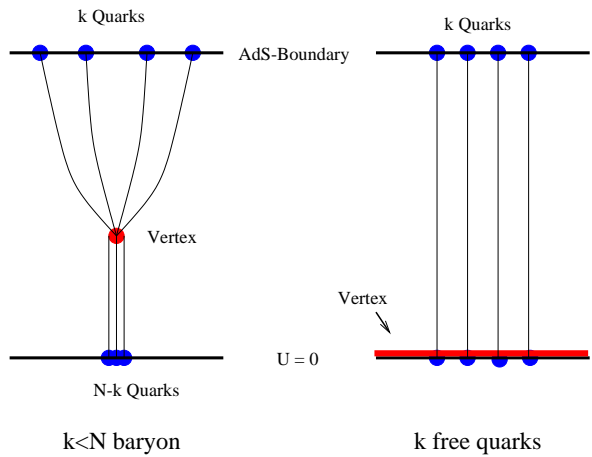}
   }
\end{center}
\caption{The $k < N$ ``baryon'' vs. $k$ free quarks.
Since the longitudinal metric vanishes at $U=0$ the the surface $U=0$
is in fact a point and hence the vertex is smeared along this
``surface'' $U=0$. As a result the string can move freely at the boundary.}
\end{figure}

\be\label{slope1}
\frac{U_x}{\sqrt{U_x^2+U^4/R^4}}= A ,~~~ \mbox{where}~~~
A=\frac{5N-4k}{4k}. 
\ee
For $k=N$ we get $A=1/4$ and for $k<N$ we have $A>1/4$.
It follows from (\ref{slope1}) that $A\leq 1$.
The upper bound , $A=1$ corresponds to $U_x\rightarrow\infty$ and $k=5N/8$.
since the strings are radial the baryon size vanishes.

The energy of the k-quarks baryon is
\be\label{ee}
E_k=\frac{U_0}{2\pi}\left[ (N-k) +N/4+k\left(
\int_1^{\infty} dy ( \frac{y^2}{\sqrt{y^4-(1-A^2)}}-1) -1\right)
\right] .
\ee
Where we have made the same kind of subtraction as in the $k=N$
configuration i.e. we have substracted the energy of $k$ quarks as
depicted in fig.4b.
For $A=1$ ($k=5N/8$) the energy vanishes which implies that the
location of the D5-brane is a moduli of the system.
For $A<1$   the energy is $b U_0$ with some negative $b$ and 
 $U_0$ is determined, as usual,  in terms of $L$.

Next we  would like to analyze   the non-supersymmetic case.
As we remarked at the begging of this section, in a confining field
theory we do not expect to find such states.
This expectation seems to be supported by the AdS supergravity
approach.
The energy of a radial string is 
\be
E=\frac{1}{2\pi}\int_{U_0}^{U_1} dU\sqrt{G_{xx}G_{uu}}\sim \log(U_0-U_T).
\ee
Therefore, the energy of a string stretched between the D5-brane and
the horizon is infinite and hence even the case $k=N-1$  cost an
infinite amount of energy.
Thus the baryonic configuration with $k=N$ is the only stable 
baryonic configuration with finite energy in
 agreement with field theory results.

{\bf Acknowledgements}
I would  like to thank J. Labastida and J. Barbon for inviting me to
present these lectures. 
The lectures are based on  work done in collaboration with 
A. Armoni, A.Brandhuber,, E.Fuchs, N.Itzhaki,
 Y. Kinar, J. Maldacena,  E. Schreiber, A Tseytlin, 
 N. Weiss  and S. Yankielowicz. I would like to thank them for the
collaboration and 
for  many useful discussions. 
This  work  is supported in part by the
US-Israrel Binational Science Foundation, by GIF - the German-Israeli
Foundation for Science Research, and by the Israel Science Foundation.  

\vskip 1 cm  

{\bf Note added}
While preparing these lecture notes the  preprint \cite{DGT}
came out. In this work the contribution of the quantum fluctuations
in the \adss case
is addressed in the framework of a Green Schwarz formulation.
It is shon there that the partition function is well defined and 
finite.  The BPS single quark, the circular loop and inifnite strip
are annalyzed.

\end{document}